\documentclass[10pt,reqno]{amsart}
\usepackage{hyperref}
\usepackage{float}
\usepackage{amscd,amssymb,amsmath,latexsym,bm}
\usepackage[mathcal,mathscr]{euscript}
\usepackage{lipsum}
\usepackage{amsfonts}
\usepackage{graphicx}
\usepackage{epstopdf}
\usepackage{algorithmic}
\usepackage{hyperref}
\usepackage{xcolor}
\usepackage{upgreek}
\usepackage{enumerate}
\usepackage{ulem}
\usepackage{gensymb}			
\usepackage[utf8]{inputenc}
\usepackage[small,nohug]{diagrams}
\diagramstyle[labelstyle=\scriptstyle]
 
\newtheorem{theorem}{Theorem}[section]

\newtheorem{proposition}[theorem]{Proposition}
\newtheorem{remark}[theorem]{Remark}
\newtheorem{definition}[theorem]{Definition}

\numberwithin{equation}{section}
\numberwithin{figure}{section}

\newcommand{\CM}{{\mathbb C}}
\newcommand{\NM}{{\mathbb N}}

\newcommand{\RM}{{\mathbb R}}

\newcommand{\ZM}{{\mathbb Z}}

\newcommand{\GM}{{\mathbb G}}

\newcommand{\Aa}{{\mathcal A}}
\newcommand{\Ee}{{\mathcal E}}
\newcommand{\Pp}{{\mathcal P}}

\newcommand{\Bb}{{\mathcal B}}

\newcommand{\Tt}{{\mathcal T}}
\newcommand{\Rr}{{\mathcal R}}

\newcommand{\Mm}{{\mathcal M}}
\newcommand{\Cc}{{\mathcal C}}

\newcommand{\Ll}{{\mathcal L}}
\newcommand{\Qq}{{\mathcal Q}}

\newcommand{\Hh}{{\mathcal H}}

\begin{document}

\title{Fermionic Topological Order on Generic Triangulations}

\author{Emil Prodan}

\address{Department of Physics and Department of Mathematical Sciences, Yeshiva University, New York, NY 10016, USA \\
\href{mailto:prodan@yu.edu}{prodan@yu.edu}}

\thanks{This work is supported by National Science Foundation through grant DMR-1823800.}


\begin{abstract} 
Consider a finite triangulation of a surface $M$ of genus $g$ and assume that spin-less fermions populate the edges of the triangulation. The quantum dynamics of such particles takes place inside the algebra of canonical anti-commutation relations (CAR). Following Kitaev's work on toric models, we identify a sub-algebra of CAR generated by elements associated to the triangles and vertices of the triangulation. We show that any Hamiltonian drawn from this sub-algebra displays topological spectral degeneracy. More precisely, if $\Pp$ is any of its spectral projections, the Booleanization of the fundmental group $\pi_1(M)$ can be embedded inside the group of invertible elements of the corner algebra $\Pp \, {\rm CAR} \, \Pp$. As a consequence, $\Pp$ decomposes in $4^g$ lower projections. Furthermore, a projective representation of $\ZM_2^{4g}$ is also explicitly constructed inside this corner algebra. Key to all these is a presentation of CAR as a crossed product with the Boolean group $(2^X,\Delta)$, where $X$ is the set of fermion sites and $\Delta$ is the symmetric difference of its sub-sets.
\end{abstract}

\maketitle


\setcounter{tocdepth}{1}

\section{Introduction and Main Statement}

A quantum many-body system is said to display topological order if it manifests topological ground state degeneracy when the underlining physical space is of non-trivial topological type and if the local excitations posses non-trivial self-statistics \cite{WenBook1}. The microscopic prototype of the topological order is the toric code introduced by Alexei Kitaev \cite{KitaevAOP2003} in the context of quantum spin-$\frac{1}{2}$ algebras. Because many calculations can be carried explicitly, this model has been instrumental for the general understanding of the topological order and of the possible applications of the latter, notably, to the topological quantum computation \cite{KitaevAOP2006,KitaevOxford2008}. The generalizations of the toric code are abundant and the body of works it spurred is very large. We invite the reader to lecture through \cite{WenBook2} for an overview and extended list of references.

\vspace{0.2cm}

The original motivation of the present work was a search for a high throughput strategy for generating many-body models that display topological degeneracy. In particular, we sought an algorithm that can be deployed on arbitrary triangulations of complex surfaces. In the process, we experimented with both spin and fermionic degrees of freedom, which in quasi 1-dimensional systems are inter-linked by the Jordan-Wigner transformation \cite[Sec.~4.1]{ColemanBook}. Recent efforts that go under the name of exact bosonization \cite{ChenAOP2018,ShuklaPRB2020}, have shown that this link can be successfully established in higher dimensions too. These efforts build on the observation made in \cite{KitaevAOP2006} that the operators:
\begin{equation}
\gamma_x = c_x^\ast + c_x, \quad s_x = c_x^\ast c_x - c_x c_x^\ast,
\end{equation}
obey commutation relations similar those occurring in the quantum spin-$\frac{1}{2}$ algebra, up to sign factors. Above, $c_x^\ast$ and $c_x$ are the fermionic creation and annihilation operators indexed by a discrete set $X$, generating the canonical anti-commutation relations (CAR) algebra CAR$(X)$. Our goal for this work, which is much more modest than that set in the exact bosonization program, is to exploit the above observation and generate toric models with fermionic degrees of freedom. Our findings are summarized in the following statement:

\begin{theorem}\label{Th:Main} Let $M$ be a compact surface of genus $g$ and $\Ll$ be one of its finite triangulations. Let $X$ be the collection of the midpoints of the edges of $\Ll$ and allow fermions to populate $X$. In a manner similar to \cite{KitaevAOP2003}, define elementary observables $\Gamma_T$ (using $\gamma$'s) and $S_V$ (using $s$'s) inside $CAR(X)$, with $T$ and $V$ being triangles and vertices of $\Ll$, respectively. Let $H$ be any self-adjoint Hamiltonian from the sub-algebra $C^\ast(\Gamma_T,S_V;\, T,V \in \Ll)$ generated by these elementary elements. Note that, since $X$ is finite, the spectrum of $H$, as an element of $CAR(X)$, consists of a finite set of eigenvalues. Then:
\begin{enumerate}[{\rm i)}] 

\item $H$ displays topological spectral degeneracy: Any spectral projection $\Pp$ corresponding to an eigenvalue can be decomposed in at least $4^g$ mutually orthogonal proper projections from the corner algebra $\Pp \, CAR(X) \, \Pp$.

\item The Booleanization $\pi_1^b(M)$ of the fundamental group $\pi_1(M)$ \footnote{In any group, the commutators form a normal sub-group and the corresponding quotient is known as the abelianization of the group. The latter can be further quoted by the sub-group generated by the second power of the abelianized elements. The result is the Booleanization of the group.} can be embedded inside the group of invertible elements from the corner algebra $\Pp \, CAR(X) \, \Pp$.

\item There is an embedding inside the corner algebra $\Pp \, CAR(X) \, \Pp$ of the non-commutative algebra:
\begin{equation}
 C^\ast(\Sigma^1,\ldots, \Sigma^{2g},\Xi^1,\ldots,\Xi^{2g}; \, \bm R),
\end{equation} 
where $\Sigma$'s and $\Xi$'s are symmetries satisfying the relations:
\begin{equation}
\bm R:\quad \Sigma^i \, \Sigma^j = \Sigma^j \,  \Sigma^i, \quad \Xi^i \, \Xi^j = \Xi^j \,  \Xi^i, \quad \Sigma^i \, \Xi^j = (-1)^{\delta_{ij}} \Xi^j \, \Sigma^i.
\end{equation}
The $\Sigma_i$'s supply the embedding mentioned at point ii).
\end{enumerate}
\end{theorem}

\begin{remark}{\rm We left out any reference to the topological excitations because the above statements show that our fermionic models reproduce exactly the same characteristics as those of the standard toric code. As such, the topological excitations can be constructed using the string operators and their self-statistics is identical to the one found in the toric code.
}$\Diamond$
\end{remark}

As one perhaps already noticed, our result is constructive and it can be deployed on any triangulation of a compact surface, regardless of its complexity. For this reason, we feel that our solution answered the original motivation we mentioned above. The starting point for our analysis is the observations that both the quantum spin-$\frac{1}{2}$ and CAR-algebras over a discrete set $X$ can be presented as crossed products with the Boolean group $(2^X,\Delta)$, whose elements are the subsets of $X$ and the binary operation is the symmetric difference.\footnote{If intersection is added as another binary operation, then $(2^X,\Delta,\cap)$ becomes a Boolean ring.} This group takes the front stage in our analysis and, in our framework, the key factor behind the topological order turns out to be a certain lattice of subgroups of $(2^X,\Delta)$.  More precisely, a certain loop sub-group $\GM_\ell$ of $(2^X,\Delta)$ can be identified which, at its turn, has a proper sub-group of local loops $\GM_\ell^{\rm loc}\simeq \langle S_V, \, V \in \Ll \rangle$, when the surface is of higher genus. Furthermore, $\GM_\ell$ can be filtered down to $\GM_\ell^{\rm loc}$ through a special lattice of sub-groups and $\GM_\ell/\GM_\ell^{\rm loc} \simeq \pi_1^b(M)$. Based on this filtration and a new ``commutant algebra argument'' stated in Proposition~\ref{Pro:CoreArg}, we are able to show $\pi_1(M)$ acts non-trivially on the spectral sub-spaces, which prompts the topological degeneracy. After presenting our arguments, it will become clear that protected spectral degeneracy can be constructed from many other sub-lattices of sub-groups. We hope that this observation can help generate additional models with protected topological degeneracy in generic crossed product algebras with the  group $(2^X,\Delta)$. 

\vspace{0.2cm}

 Below, we use pointed remarks to provide more insight into our result and to establish relations with other existing results:

\begin{remark}{\rm The algebra $C^\ast(\Gamma_T,S_V;\, T,V \in \Ll)$ is non-commutative as opposed to the corresponding algebra of the classic toric code, generated by the $A$'s and $B$'s  which commute. The non-commutativity spurs from the $\gamma$'s, which anti-commute over long ranges. Hence, the non-commutative character encountered in our work is different from the non-abelian generalization of the toric code, such as those based on Drinfeld's quantum double (see the second half of \cite{KitaevAOP2003}). It is, however, important for our arguments that $\Gamma_T$'s and $S_V$'s commute.
}$\Diamond$
\end{remark}

\begin{remark}{\rm In the context of topological quantum computing, the corner algebra $\Pp CAR(X) \Pp$ can be still considered as the algebra of protected physical observables. The dynamics $A \mapsto e^{\imath t H} A e^{-\imath t H}$ of these observables under $H$ is suppressed and these observables can be acted on by adiabatic braiding of the anyons.
}$\Diamond$
\end{remark} 

\begin{remark}{\rm The point iii) can be reformulated by saying that  the group $\ZM_2^{\times 4g}$ accepts a projective representation inside the multiplicative group of the unitary elements of $\Pp CAR(X) \Pp$.
}$\Diamond$
\end{remark}

\begin{remark}{\rm Note that the statement about topological degeneracy comes before the one about the representation of the non-commutative algebra. This indicates that we found a new proof of the topological degeneracy, which does not rely on the third statement. For example, in \cite{KitaevAOP2003}, the proof of the ground state degeneracy relies entirely on the existence of a projective representation of $\ZM_2^{\times 4g}$. Our proof rather highlights the embedding of $\pi_1^b(M)\simeq \ZM_2^{2g}$ into the algebra of protected physical observables.
}$\Diamond$
\end{remark}

\begin{remark}{\rm As opposed to other existing works on the fermionic topological order, {\it e.g.} \cite{GuPRB2014}, we will not make any use of the relation between the quantum spin-$\frac{1}{2}$ and CAR algebras. Let us point out that in the thermodynamic limit, the quantum spin-$\frac{1}{2}$ algebra and the CAR algebra are not isomorphic anymore, but can be embedded into a larger algebra such that their even sectors coincide \cite{BratelliBook2}[pp.~258]. This can create problems when taking the thermodynamic limit of fermionic models inspired from spin models.
}$\Diamond$
\end{remark}

\begin{remark}{\rm We feel that the crossed product presentation of the CAR-algebra is very close in spirit to Renault's grupoid presentation \cite[pp. 129]{RenaultBook}. Perhaps, our entire analysis can be carried within that formalism, which is something we are presently looking into.
}$\Diamond$
\end{remark} 

\section{Algebra of Canonical Anti-Commutation Relations (CAR)}

We collect in this section the most basic facts about CAR-algebra, which, in essence, is one of the simplest and least interesting $C^\ast$-algebra. Many of the statements follow from the general literature on $C^\ast$-algebra, as adopted to the particular context of CAR-algebras. We will exclusively deal with CAR-algebras over finite dimensional Hilbert spaces.

\subsection{Definition and standard presentation} We consider the algebra of canonical anti-commutation relations over a discrete finite set $X$ of points, denoted by $CAR(X)$. It is generated by the elements $c_x$, $x\in X$, their conjugates and by a unity, all subjected to the relations:
\begin{equation}\label{Eq:CAR1}
\Rr: \ \ c_x c_{x'} + c_{x'} c_x=0,  \quad c_x^\ast c_{x'} + c_{x'} c_x^\ast = \delta_{x x'}\, 1,
\end{equation}
for all $x,x'\in X$. One will write in short:
\begin{equation}
CAR(X) = C^\ast(c_x, c_x^\ast, \, x \in X; \, \Rr),
\end{equation}
a notation which will be used quite often in these notes. 

\vspace{0.2cm}

The elements of $CAR(X)$ will be denoted with capital letter $A$, $B$, etc.. A generic element of $CAR(X)$ can be presented as:
\begin{equation}\label{Eq:GenericElement}
A = \sum_{J,J' \subseteq X} a_{J,J'} \prod_{x \in J} c_x^\dagger \prod_{x' \in J'} c_{x'},
\end{equation}
where the coefficients $a_{J,J'} \in \CM$ are uniquely defined up to a sign. To fix the signs, we order the set $X$ once and for all and equip every subset, like $J$ and $J'$ above, with the order induced from $X$. With these conventions, the sum in \eqref{Eq:GenericElement} is over ordered sets and the products $\prod_{x \in J} c_x^\dagger$ and $\prod_{x' \in J'} c_{x'}$ are ordered accordingly. As such, the signs of the coefficients $a_{J,J'}$ are fully determined by the notation. We should also specify that the empty set is a subset of $X$, hence the sum in \eqref{Eq:GenericElement} includes $J=\emptyset$ or $J'=\emptyset$. In particular, the term corresponding to $J=J'=\emptyset$ is simply $a_{\emptyset,\emptyset}\, 1$.

\vspace{0.2cm}

Any self-adjoint element:
\begin{equation}
H = \sum_{J,J' \subseteq X} h_{J,J'} \prod_{x \in J} c_x^\dagger \prod_{x' \in J'} c_{x'}, \quad H=H^\ast,
\end{equation} 
from $CAR(X)$ can serve as the generator of the dynamics of the physical observables:
\begin{equation}
CAR(X) \ni A \mapsto A(t) = e^{\imath t H} A e^{-\imath t H}, \quad t \in \RM.
\end{equation}
One should be aware that this applies only for finite $X$. If $X$ is infinite, then CAR-algebra is defined as a directed limit and the generator of the time-evolution generally falls outside of CAR in this limit.\footnote{This is the major difficulty for the operator-theoretic topological classification.} Since we will avoid any representation of CAR on Hilbert spaces, let us recall that, in the pure algebraic setting, the resolvent set of $H$ consists of all the points $z$ in the complex plane for which $H-z$ belongs to the group ${\rm GL}\big (CAR(X)\big)$ of invertible elements of $CAR(X)$. The spectrum of $H$ is the complement of the resolvent set and, since $X$ is finite, this spectrum consists of a finite set of eigenvalues $\lambda_1$, $\lambda_2$, \ldots, located on the real axis. The spectral projections are defined by $\Pi_j H = H \Pi_j = \lambda_j \Pi_j$ and can be computed as $\Pi_j=P_j(H)$, where $P_j$'s are the so called interpolating polynomials, uniquely defined by $P_j(\lambda_i) = \delta_{ij}$. We mentioned this detail only because it is now straightforward to see that, if $H$ belongs to a sub-algebra of $CAR(X)$, then so do all spectral projections of $H$.

\vspace{0.2cm}

 Inside $CAR(X)$, there are the following familiar projections, which we write out explicitly in order to fix the notation:
\begin{equation}
n_x = c_x^\ast c_x, \quad n_x^2 = n_x^\ast = n_x, \quad x \in X,
\end{equation}
and their complements: 
\begin{equation}
n_x^\bot = 1 - n_x = c_x c_x^\ast, \quad x \in X,
\end{equation} 
which all commute with each other. In fact $n_x$ and $n_x^\bot$ commute with any element $A$ for which $x$ is not part of any of the $J$ and $J'$ sets in the expression \eqref{Eq:GenericElement}. 

\subsection{The unique trace state}

$CAR(X)$ accepts a unique normalized trace, that is, a linear map $\Tt$ from itself to $\CM$ such that:
\begin{equation}
\Tt(AB)=\Tt(BA), \quad \forall \, A,B \in CAR(X),
\end{equation} 
and normalized as $\Tt(1)=1$. This can be seen by observing that any trace returns zero when evaluated on the commutator space:
\begin{equation}
[CAR(X),CAR(X)] : =  \CM{\rm -Span}\big \{ [A,B], \ A,B \in CAR(X) \big \}.
\end{equation}
Using \eqref{Eq:CAR1}, in particular:
\begin{equation}
c_x = [c_x,n_x], \quad c_x^\ast = [n_x,c_x^\ast], \quad n_x - n_x^\bot = [c_x^\ast, c_x],
\end{equation}
 one can convince oneself that the commutator space is quite large, in particular that:
\begin{equation}
CAR(X)/[CAR(X),CAR(X)] = \CM \cdot 1.
\end{equation}
Since any trace on $CAR(X)$ factors through the canonical projection \cite[Lemma 1.36]{EmmanouilBook}:
\begin{equation}\label{Eq:Chi}
\chi : CAR(X) \rightarrow CAR(X)/[CAR(X),CAR(X)],
\end{equation}
there is indeed unique normalized trace on $CAR(X)$. In particular, for $n_x$ and $n_x^\bot$ elements, this trace returns $\Tt(n_x) =\Tt(n_x^\bot) = \tfrac{1}{2}$. Another way to see this is to use the explicit isomorphism between $CAR(X)$ and the full matrix algebra $M_{2^{|X|}}(\CM)$ \cite{DavidsonBook}. Throughout, $|\cdot|$ will represent the cardinal of a set.

\begin{remark}{\rm In the presentation \eqref{Eq:GenericElement}, it is not always straightforward to evaluate the trace, as the only generic algorithm relies on evaluating the map $\chi$ from \eqref{Eq:Chi} on a particular element. The following section introduces a new presentation of $CAR(X)$ in which evaluation of the trace is a trivial process.
}$\Diamond$
\end{remark}

\vspace{0.2cm}

Given the last statement, the structure of the projections in $CAR(X)$ is extremely simple: The range of the trace on projections, i.e. of the dimension function, takes any value from the set $\{n\cdot 2^{-|X|}, \, n \in \NM, \, n \leq 2^{|X|} \}$. Hence, the trace of the minimal projections is $\Tt(P_m) = 2^{-|X|}$. Two projection of same trace can be connected by a similarity tranformation, $\Tt(P)=\Tt(Q) \Rightarrow P= UQP$, for some unitary element $U$. Furthermore, since $CAR(X) \simeq M_{2^{|X|}}(\CM)$ has stable rank 1, two similar projections can be homotopically connected \cite{Giol}.  

\vspace{0.2cm}

A gapped Hamiltonian from $CAR(X)$ is a pair $(H,G)$, where $H$ is self-adjoint and $G$ is a connected component of the resolvent set $\RM \setminus {\rm Spec}(H)$ ($G$ also symbolizes the mid point of the gap). Recall that any gapped Hamiltonian from $CAR(X)$ can be reduced to a symmetry by spectral flattening $H \rightarrow {\rm sign}(H-G)$. Since a symmetry is of the form $1-2P$ with $P$ a projection, classifying gapped Hamiltonians without symmetry constraints is same as classifying projections. According to the previous paragraph, if the classification is by homotopy, as usually done in the physics literature, then all topological invariants that can be associated to gapped Hamiltonians from $CAR(X)$ are derived from the dimensions of their gap projections. The lesson is that nothing interesting is to be expected if the models are from and the homotopy deformations are in the whole $CAR(X)$. In order to see something topologically interesting, one needs to restrict to sub-algebras. The text of Theorem~\ref{Th:Main} definitely convey that.

\subsection{States and GNS representations}

As we already mentioned, we will avoid any explicit representation of the $CAR$ algebra. However, many of us are very familiar with the Fock representation of $CAR(X)$, hence we include here a few comments about it.

\vspace{0.2cm}

The unique trace $\Tt$ is positive definite, {\it i.e.} it returns strictly positive values on the elements of the form $A^\ast A$, $A \neq 0$. As such, $\Tt$  can be used to define quantum states. For example, if $F$ is a positive element, then:
\begin{equation}\label{Eq:OmegaState}
\omega_F(A) : = \Tt(AF) / \Tt(F)
\end{equation}
defines a state on $CAR(X)$. Ground states come from projections $P=P^2=P^\ast$.

\vspace{0.2cm} 

Associated to any state $\omega$, there is a standard GNS Hilbert space $\Hh_\omega$, supplied by the completion of the linear structure of $CAR(X)$ w.r.t. the norm coming from the scalar product:
\begin{equation}
\langle A,B \rangle_\omega = \omega(A^\ast B), \quad A,B \in CAR(X).
\end{equation}
One usually denotes by $|B\rangle_\omega$ the class of $B \in CAR(X)$ in that completion. The elements of $CAR(X)$ are represented on $\Hh_\omega$ as bounded operators on this Hilbert space via \cite{DixmierBook1}:
\begin{equation}
\pi_\omega(A) |B\rangle_\omega = |A B \rangle_\omega, \quad \forall \ A,B \in CAR(X).
\end{equation}
The familiar Fock-space representation can be obtained as a particular GNS representation, as explained below. Let us mention one more thing, that the group of invertible elements $V \in {\rm GL}\big (CAR(X)\big )$ with $\omega \circ {\rm Ad}_V=\omega$ (${\rm Ad}_V=$ adjoint action of $V$) accepts a right-acted representation:
\begin{equation}
\pi'_\omega(V) |B\rangle_\omega = |B V^{-1}\rangle, \quad V\in {\rm GL}\big (CAR(X)\big ).
\end{equation}
These are linear operators that all commute with every single $\pi_\omega(A)$. 

\vspace{0.2cm}

Going back to the Fock representation, let:
\begin{equation}\label{Eq:NElement}
N = \sum_{x \in X} c_x^\dagger c_x = \sum_{x \in X} n_x \in CAR(X),
\end{equation}
be the physical observable corresponding to the fermion number and consider the following partition of unity:
\begin{equation}
1 = \prod_{x \in X} (n_x + n_x^\bot) = \sum_{J \subseteq X} P_J, \quad P_J = \prod_{x \in J} n_x \prod_{y \in X \setminus J} n_y^\bot.
\end{equation}
Then $P_J P_{J'} = \delta_{J \, J'}$ and:
\begin{equation}
N = \sum_{J \subseteq X} |J| \, P_J,
\end{equation}
where $|\cdot|$ denotes the cardinal of a set. In particular, the spectrum of $N$ consists of $\{0,1,\ldots,|X| \}$. Furthermore, since $\Tt(P_J) = 2^{-|X|}$, it follows that $\Tt(N) = \tfrac{1}{2}|X|$, which also follows directly from \eqref{Eq:NElement}. The projection $P_\emptyset$ corresponding to the empty set is special: $N P_\emptyset =0$. The standard Fock representation of $CAR(X)$ is the GNS representation corresponding to the state:
\begin{equation}
CAR(X) \ni A \mapsto \omega_\emptyset(A) = 2^{|X|} \,  \Tt(A P_\emptyset) \in \CM.
\end{equation}
This state can be characterized more directly as the unique state in which the monomians $M_{JJ'}=\prod_{x \in J}c^\ast_x \prod_{x' \in J'} c_{x'}$ satisfy:
\begin{equation}
\omega_\emptyset\big (M_{JJ'}^\ast M_{\tilde J \tilde J'} \big ) = \delta_{J\tilde J} \delta_{J'\cup\tilde J',\emptyset}, \quad J,J',\tilde J, \tilde J' \subseteq X.
\end{equation}
The vector $|1\rangle_\emptyset$ of the GNS Hilbert space corresponding to the unit element stands for the Fock vacuum $|\emptyset \rangle$, $\pi_\emptyset(N)$ becomes the particle number operator and $\pi_\emptyset (c_x)$ and $\pi_{\emptyset}(c_x^\ast)$ become lowering (annihilation) and raising (creation) operators for $\pi_\emptyset(N)$, respectively.

\section{CAR as a Crossed Product Algebra}

We supply here a presentation of the CAR-algebra that will enable us to view $CAR(X)$ as a crossed product between a Clifford algebra and the Boolean group $(2^X,\Delta)$. 

\subsection{Elementary self-adjoint generators}\label{Sec:ElementaryGen}

Consider the self-adjoint elements from $CAR(X)$: 
\begin{equation}
\tilde \gamma_x = c_x + c_x^\ast, \quad \tilde s_x = n_x - n_x^\bot, \quad x \in X.
\end{equation}
We show below that $CAR(X)$ can be presented in terms of $\tilde \gamma$ and $\tilde s$ elements.

\begin{proposition} We have:
\begin{equation}
CAR(X) \simeq  C^\ast(\gamma_x,\, s_x, \ x \in X; \ \Rr'),
\end{equation}
where the relations $\Rr'$ are:
\begin{equation}\label{Eq:RPS}
\Rr'_1: \ \ s_x^2 =1, \ \ s_x s_{x'} = s_{x'} s_x, \quad \forall \ x,x' \in X;
\end{equation}
\begin{equation}\label{Eq:RPG}
\Rr'_2: \ \ \gamma_x \gamma_{x'} + \gamma_{x'} \gamma_x = 2 \delta_{x,x'} \, 1, \quad \forall \ x,x' \in X; 
\end{equation}
\begin{equation}\label{Eq:RPGS}
\Rr'_3: \ \ s_x \gamma_{x'} s_x = (-1)^{\delta_{x,x'}} \gamma_{x'}, \quad \forall \ x,x' \in X.
\end{equation}
\end{proposition}

\noindent {\bf Proof.} Using the commutation relations $\Rr$ and $\Rr'$, we checked that mapping:
\begin{equation}
CAR(X) \ni c_x \mapsto \tfrac{1}{2}\gamma_x(1+s_x) \in C^\ast(\gamma_x,s_x, \, x \in X)
\end{equation}
extends to a unique $\ast$-homomorphism of algebras. Furthermore, using the relations $\Rr'$, we have:
\begin{equation}\label{Eq:I1}
\gamma_x = \tfrac{1}{2} \gamma_x(1+s_x) + \tfrac{1}{2}\big ( \gamma_x(1+s_x) \big )^\ast, 
\end{equation}
and:
\begin{equation}\label{Eq:I2}
s_x = \Big ( \tfrac{1}{2}\gamma_x(1+s_x) \Big )^\ast \,  \Big (\tfrac{1}{2}\gamma_x(1+s_x) \Big )-\Big (\tfrac{1}{2}\gamma_x (1+s_x) \Big)\, \Big ( \tfrac{1}{2}\gamma_x(1+s_x) \Big )^\ast.
\end{equation}
Then, using the $\Rr$ and $\Rr'$ relations again,  we checked that the mapping:
\begin{equation}
\gamma_x \mapsto c_x^\ast +c_x, \quad s_x \mapsto c^\ast_x c_x - c_x c^\ast_x, 
\end{equation}
extends to a unique $\ast$-homomorphism of algebras and, from \eqref{Eq:I1} and \eqref{Eq:I2}, one can immediately see that the two homomorphisms are inverse to each other. $\square$

\begin{remark}{\rm In several places, it will be useful to reformulate the commutation relations for $\gamma$'s as:
\begin{equation}
\gamma_x^2 =1, \quad \gamma_x \gamma_{x'} \gamma_x = -(-1)^{\delta_{xx'}} \gamma_{x'}, \quad \forall \ x,x' \in X.
\end{equation}
}
\end{remark}

\subsection{Presentation as a crossed product}

According to relations $\Rr'$, the $s$-elements are commuting symmetries and they form a multiplicative abelian group $\bm S(X)$. On the other hand, the $\gamma$-elements form a Clifford algebra $\bm \Gamma(X)$,  which is invariant under conjugation with elements from $\bm S(X)$. This tells us that $CAR(X)$ can be viewed as the crossed product $\bm \Gamma(X) \rtimes_{\rm Ad} \bm S(X)$ with the action:
\begin{equation}
\bm S(X) \ni S \mapsto {\rm Ad}_S \in {\rm Aut}\big (\bm \Gamma(X) \big ), \quad {\rm Ad}_S(\Gamma)= S \Gamma S^{-1}.
\end{equation} 
As such, any element from $CAR(X)$ can be presented uniquely as:
\begin{equation}\label{Eq:Presentation2}
A = \sum_{S \in \bm S(X)} \Gamma(S) \, S, \quad \Gamma(S) \in \bm \Gamma(X),
\end{equation}
and multiplication takes the form:
\begin{equation}
A A' = \sum_{S \in \bm S(X)} \Big ( \sum_{\tilde S \in \bm S(X)} \, \Gamma(\tilde S) \, {\rm Ad}_{\tilde S}\big (\Gamma'(\tilde S^{-1} S)\big ) \Big ) \, S.
\end{equation}
Furthermore, any element of $\bm \Gamma(X)$ takes the form:
\begin{equation}\label{Eq:GammaElement}
\Gamma = \sum_{J \subseteq X} g_J \, \prod_{x \in J} \gamma_x \equiv \sum_{J \subseteq X} g_J \Gamma_J,
\end{equation}
with the coefficients $g_J$ determined uniquely. Let us recall that the subsets $J$ are ordered and the products $\prod_{x \in J} \gamma_x$ are also ordered accordingly. Also, the empty set is a subset of $X$ and the term of the sum corresponding to $J=\emptyset$ is simply $g_\emptyset \, 1$.

\vspace{0.2cm}

\begin{remark}\label{Re:CP}{\rm The new presentation has several immediate advantages, which can be useful in the generic investigations of the CAR-algebras:
\begin{itemize}

\item First, the structure of the idempotents of finite and infinite Clifford algebras is known explicitly \cite{Wene}. 

\item The trace can be easily evaluated on generic elements. Indeed, with the notations from \eqref{Eq:Presentation2} and \eqref{Eq:GammaElement}:
\begin{equation}
\Tt(A) = \Tt(\Gamma_1), \quad \Tt(\Gamma) = g_\emptyset.
\end{equation}

\item The monomials are mutually orthogonal and normalized:
\begin{equation}
\Tt\Big ( (\Gamma_J S_{J'})^\ast (\Gamma_{\tilde J} S_{\tilde J'}) \Big) = \delta_{J\tilde J} \, \delta_{J'\tilde J'}.
\end{equation}

\item The presentation brings the CAR-algebra closer to that of quantum spin algebras, hence inspiration can be drawn from a large body of work on the latter. $\Diamond$

\end{itemize}
}
\end{remark}

\begin{proposition} $\bm S(X) \simeq (2^X,\Delta)$.
\end{proposition}
\noindent{\bf Proof.} Any element of $\bm S(X)$ is of the form:
\begin{equation}\label{Eq:SJ}
S_J = \prod_{x \in J} s_j, \quad J \subseteq X,
\end{equation}
and:
\begin{equation}
S_J \, S_{J'} = \Big ( \prod_{x \in J\setminus J'} s_x \Big ) \Big ( \prod_{x \in J \cap J'} s_x^2 \Big ) \Big ( \prod_{x \in J' \setminus J} s_x \Big ),
\end{equation}
which leads to:
\begin{equation}\label{Eq:SMult}
S_J \, S_{J'} = S_{J \Delta J'}, \quad J \Delta J' = (J\setminus J') \cup (J' \setminus J).
\end{equation}
Hence:
\begin{equation}\label{Eq:MainIso}
2^X \ni J \mapsto S_J = \prod_{x \in J} s_x \in \bm S(X)
\end{equation} 
supplies a group morphism, which is clearly bijective. $\square$

\vspace{0.2cm}

The conclusion is that the $CAR$ algebra can be identified with the crossed product algebra $\bm \Gamma(X) \rtimes (2^X,\Delta)$.

\begin{remark}{\rm For an infinite directed set $X$, both $\bm \Gamma(X)$ and the group $\bm S(X)$ can be defined as directed limits. It should be relatively easy to verify if the directed limit of the crossed product coincides with the standard definition of the CAR algebra on infinite sets.
}$\Diamond$
\end{remark}

\subsection{Structure of the new presentation}
\label{Sec:Rules}

There is a whole lot more structure in the new presentation. The monomials of the Clifford algebra $\bm \Gamma(X)$ also have a relation with the group $(2^X,\Delta)$:
\begin{equation}\label{Eq:GammaJ}
\Gamma_J = \prod_{x \in J} \gamma_x, \quad \Gamma_J \Gamma_{J'} = (-1)^{\eta_{JJ'}} \, \Gamma_{J \Delta J'}, \quad J,J' \subseteq X.
\end{equation} 
In general, the sign factor depends on $J$ and $J'$ as well as on the order assigned to $X$. The associativity of the multiplication in $CAR(X)$ forces $\eta$ to be a two-cocycle and the monomials of $\bm \Gamma(X)$ supply a projective representation of the Boolean group $(2^X, \Delta)$. Furthermore:
\begin{equation}\label{Eq:Gamma0}
\Gamma_J \Gamma_{J'} = (-1)^{|J| \cdot |J'|-|J \cap J'|} \, \Gamma_{J'} \Gamma_J, \quad J,J' \subseteq X.
\end{equation}
The commutation relations between $S$ and $\Gamma$ monoidals are quite simple:
\begin{equation}\label{Eq:CommGS}
\Gamma_J S_{J'} = (-1)^{|J \cap J'|} S_{J'} \Gamma_J, \quad J,J' \subseteq X.
\end{equation}
In particular, they commute whenever $|J \cap J'|$ is even.

\vspace{0.2cm}

Based on all the above, an element of $CAR$-algebra can be uniquely presented as:
\begin{equation}\label{Eq:AEl}
A = \sum_{J,J' \subseteq X} a_{J,J'} \Gamma_J S_{J'}, \quad a_{J,J'} \in \CM.
\end{equation}
The ``Fourier'' coefficients $a_{J,J'}$ can be extracted by using the trace:
\begin{equation}
a_{J,J'}= \Tt \Big (( \Gamma_J S_{J'})^\ast A \Big), \quad J,J' \subseteq X.
\end{equation}
Furthermore, we have the following simple rules of multiplication between the $S$-monomials and elements presented as in \eqref{Eq:AEl}:
\begin{align}\label{Eq:MRule1}
S_L A & = \sum_{J,J' \subseteq X} (-1)^{|J \cap L|}  a_{J,J'} \Gamma_J S_{L \Delta J'}, \\ 
A S_L & = \sum_{J,J' \subseteq X}  a_{J,J'} \Gamma_J S_{L \Delta J'}.\label{Eq:MRule11}
\end{align}
For multiplication by $\Gamma$-monomials, the rules are:
\begin{align}\label{Eq:MRule2}
\Gamma_L A & = \sum_{J,J' \subseteq X} (-1)^{\eta_{LJ}}  a_{J,J'} \Gamma_{L \Delta J} S_{J'}, \\ \label{Eq:MRule22}
A \Gamma_L & = \sum_{J,J' \subseteq X} (-1)^{\eta_{JL}+|L \cap J'|} a_{J,J'} \Gamma_{J \Delta L} S_{J'}.
\end{align}

\begin{remark}\label{Re:ZeroA}{\rm Since the following facts will play a certain role in our future arguments, we want to state them explicitly, even though they may appear obvious:
\begin{equation}
\sum_{J,J' \subseteq X} a_{J,J'} \Gamma_J S_{J'} = \sum_{J,J' \subseteq X} a'_{J,J'} \Gamma_J S_{J'}
\end{equation} 
if and only if $a_{J,J'} = a'_{J,J'}$ for all $J,J' \subseteq X$. Likewise, for the presentation \eqref{Eq:Presentation2}:
\begin{equation}
\sum_{S \in \bm S(X)} \Gamma_S S = \sum_{S \in \bm S(X)} \Gamma'_S S
\end{equation}
if and only if $\Gamma_S = \Gamma'_S$ for all $S \in \bm S(X)$.
}$\Diamond$
\end{remark}

\section{Elementary Operators of Triangulations}

\subsection{Why triangulations?}

Every compact orientable surface admits a finite triangulation \cite{CarforaBook} and, since the theme of these notes is quantum degeneracy on higher genus surfaces, we feel that it is imperative to develop our models and analysis in a manner that can be adapted and applied to any generic triangulation. Let us point out that, in \cite{KitaevAOP2003}, Kitaev works with arbitrary lattices, a far more generic setting than ours. We should warn the reader, however, that some of our arguments leading to the topological degeneracy fail if the lattice is not a triangulation.

\begin{definition}{\rm A triangulation $\Ll$ of a surface $M$ is a simplicial complex that is homeomorphic to $M$ and such that each edge of the simplex belongs to a 3-leg cycle and never to a 2-leg cycle. The 3-leg cycles are referred to as the elementary triangles.
}
\end{definition}

\begin{remark}{\rm It is important to mention that we only consider surfaces without boundaries. In these cases, any edge belongs to two and only two adjacent triangles. This is an important detail.
}$\Diamond$
\end{remark} 

\vspace{0.2cm}

\begin{figure}
\center
\includegraphics[width=0.9\linewidth]{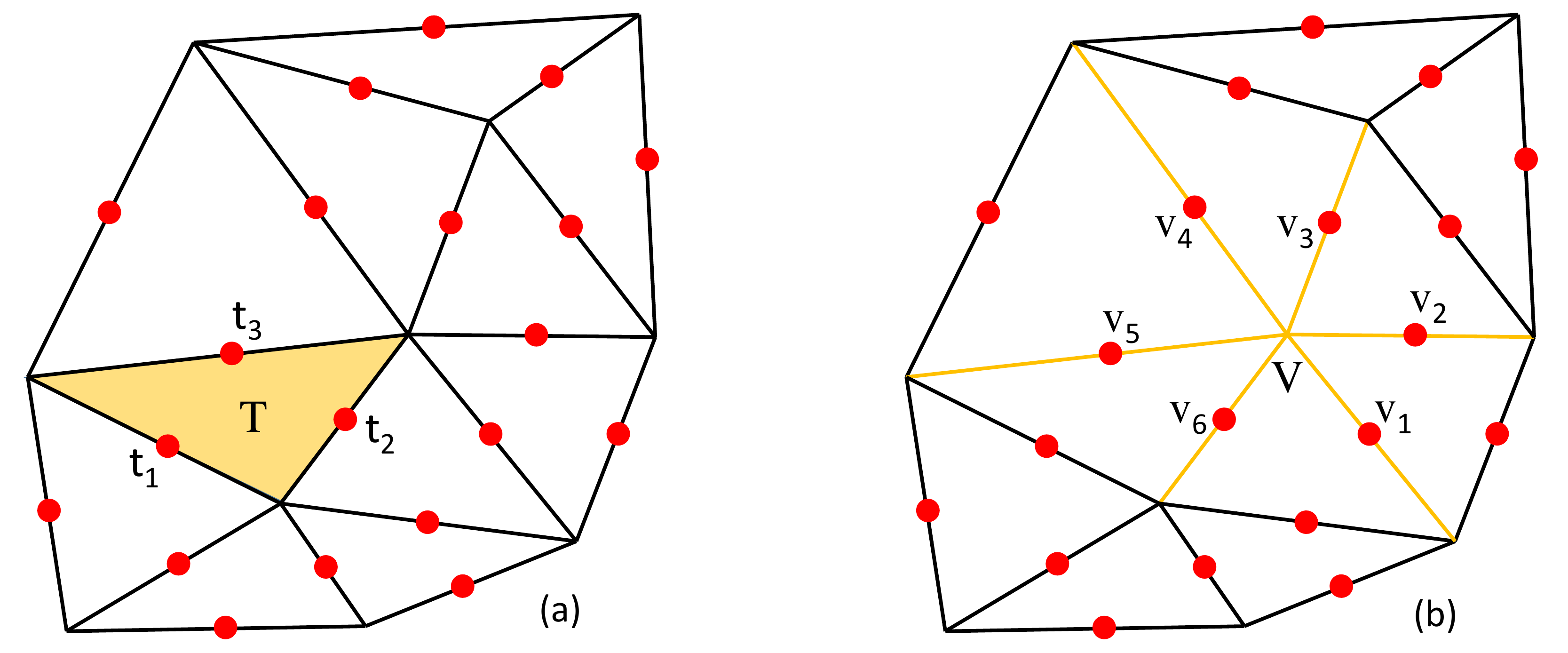}
\caption{\small Definition of elementary elements on a triangulation $\Ll$, shown here only partially. The mid-points of the edges, indicated by red dots, define the sites that can be populated with fermions. (a)  Elementary triangles of $\Ll$, such as $T = {\rm Order}\{t_1,t_2,t_3\}$, support the triangle elements $\Gamma_T = \prod_{t \in T} \gamma_t$. (b) The vertices of $\Ll$, such as $V={\rm Order}\{v_1,\ldots,v_6\}$, support the vertex elemements $S_V = \prod_{v \in V} s_v$. Here, Order =  ordering w.r.t. the order assigned to $\Ll$.}
\label{Fig:Triangulation1}
\end{figure}

Fig.~\ref{Fig:Triangulation1} illustrates a section of a generic triangulation $\Ll$. Sites are placed on the edges of the triangulation and it is at these sites where fermions reside. Hence, the set $X$ of the previous section is the reunion of these sites, which will be endowed with an order. The symbol $\Ll$ will be thought as incorporating all the information about the triangulation, such as the edges, vertices, fermion sites, order of $X$, etc.. The elementary triangles of $\Ll$ will be identified with their three edges, such as $T=\{t_1,t_2,t_3\}$ in Fig.~\ref{Fig:Triangulation1}. The set $T$ will be ordered according to the order pre-assigned to $X$. Likewise, a vertex $V$ will be identified with its legs, such as $V=\{v_1, \ldots,v_6\}$ in Fig.~\ref{Fig:Triangulation1}. This set will also be ordered.

\subsection{The elementary triangle elements}

We will consider the monomials $\Gamma_J$ defined in \eqref{Eq:GammaJ} with $J$'s restricted to the elementary triangles of $\Ll$:
\begin{equation}
\Gamma_T = \prod_{t\in T} \gamma_t, \quad T \in \Ll.
\end{equation}
Recall that $T$, as a set, and the product over $T$ are both ordered. We will denote the sub-algebra of $CAR(X)$ generated by $\Gamma_T$'s as:
\begin{equation}
\bm \Gamma_T=C^\ast(\Gamma_T, \, T \in \Ll) \subset CAR(X).
\end{equation}
From \eqref{Eq:GammaJ}, we know that:
\begin{equation}\label{Eq:ProjectiveG}
\Gamma_T \Gamma_{T'} = (-1)^{\eta_{TT'}} \Gamma_{T \Delta T'}, \quad T,T' \in \Ll.
\end{equation}
It follows that all $\Gamma_T^2$ are proportional to the identity and by multiplying them with the phase factor $\imath^{\eta_{TT}}$, all $\Gamma_T$'s can be made into symmetries:
\begin{equation}\label{Eq:GammaSymmetry}
(\imath^{\eta_{TT}}\Gamma_T)^2 = 1, \quad (\imath^{\eta_{TT}}\Gamma_T)^\ast = \imath^{\eta_{TT}}\Gamma_T.
\end{equation} 
 Also, $\Gamma_T$'s commute with each other when $T$ and $T'$ share an edge but they anti-commute when $T$ and $T'$ share no edge. For this reason, the $\Gamma_T$'s cannot be regarded as stabilizer elements, as in the toric code. This, is a major difference between the fermionic and spin models yet, quite surprisingly, it has little impact on our conclusions about the topological degeneracy of the ground states.

\subsection{The elementary vertex operators}

The vertex operators are defined by the monomials $S_J$ defined in \eqref{Eq:SJ}, with $J$'s restricted to the vertices of $\Ll$:
\begin{equation}\label{Eq:AV}
S_V = \prod_{v \in V} s_v, \quad V \in \Ll.
\end{equation}
Since the $s$ observables commute with each other, the ordering of $s_v$'s pose no problem in \eqref{Eq:AV} and clearly $S_V$'s commute with each other:
\begin{equation}\label{Eq:AComm}
S_V S_{V'} = S_{V'} S_V, \quad \forall \ V,V' \in \Ll.
\end{equation}
Obviously, $S_V$'s are elements of the abelian group $\bm S(X)$ but, nevertheless, let us state explicitly that the vertex elements are symmetries:
\begin{equation}
S_V ^2 = 1, \quad S_V^\ast = S_V, \quad \forall \ V \in \Ll.
\end{equation}
Further formalization of the vertex elements will be supplied in the next section.

\subsection{Inter-commutation relations}

We now discuss the relation between triangle and vertex observables. Since the triangle and vertices can share zero or two edges, $|T \cap V| = \rm{even}$ and it follows directly from \eqref{Eq:MRule2} and \eqref{Eq:MRule22} that:
\begin{equation}\label{Eq:BA}
\Gamma_T S_V = S_V \Gamma_T, \quad \forall \ T,V \in \Ll,
\end{equation}
and this, together with \eqref{Eq:AComm}, lists all commutations available for the elementary elements of a triangulation.

\begin{remark}{\rm $\bm \Gamma_T$ can be identified with the fixed point sub-algebra of $\bm \Gamma(X)$ with respect to all conjugations by $S_V$'s. Since: 
\begin{equation}
C^\ast(\Gamma_T,S_V; \, T,V \in \Ll)=C^\ast(\bm \Gamma_T,S_V; \, V \in \Ll),
\end{equation}
it becomes clear that this sub-algebra is fully determined by the $S_V$'s. This observation may simplify the task of defining the analog of the sub-algebra $C^\ast(\Gamma_T,S_V; \, T,V \in \Ll)$ in other contexts, such as different types of lattices. Furthermore, our argument leading to the topological spectral degeneracy does not require the explicit computation of this fixed point sub-algebra.
}$\Diamond$
\end{remark}

Compared with the toric code, one set of commutation relations is missing and, at first sight, it may seem that the loss of commutation between the $\Gamma_T$ observables will put the argument from \cite{KitaevAOP2003} in jeopardy. As we shall see, however, the spectral degeneracy steams entirely from the vertex observables. 

\section{The loop sub-group of $(2^X,\Delta)$}

In this section, we place the elementary vertex elements in their proper context.

\subsection{Admissible contours defined}

Taking cues from Kitaev's constructions \cite{KitaevAOP2003}, we introduce a special set of contours:

\begin{definition}{\rm The admissible contours are defined by the rules:
\begin{itemize}

\item A contour skips in straight manner from one fermion site to another.

\item Each leg of a contour joins two edges of same triangle. In other words, each leg is contained in one and only one triangle of $\Ll$.

\item The contour, as it meanders over the triangulation, intersects either none or precisely two sides of any triangle of $\Ll$.

\end{itemize} 
}
\end{definition}

 \begin{figure}
\center
\includegraphics[width=0.8\linewidth]{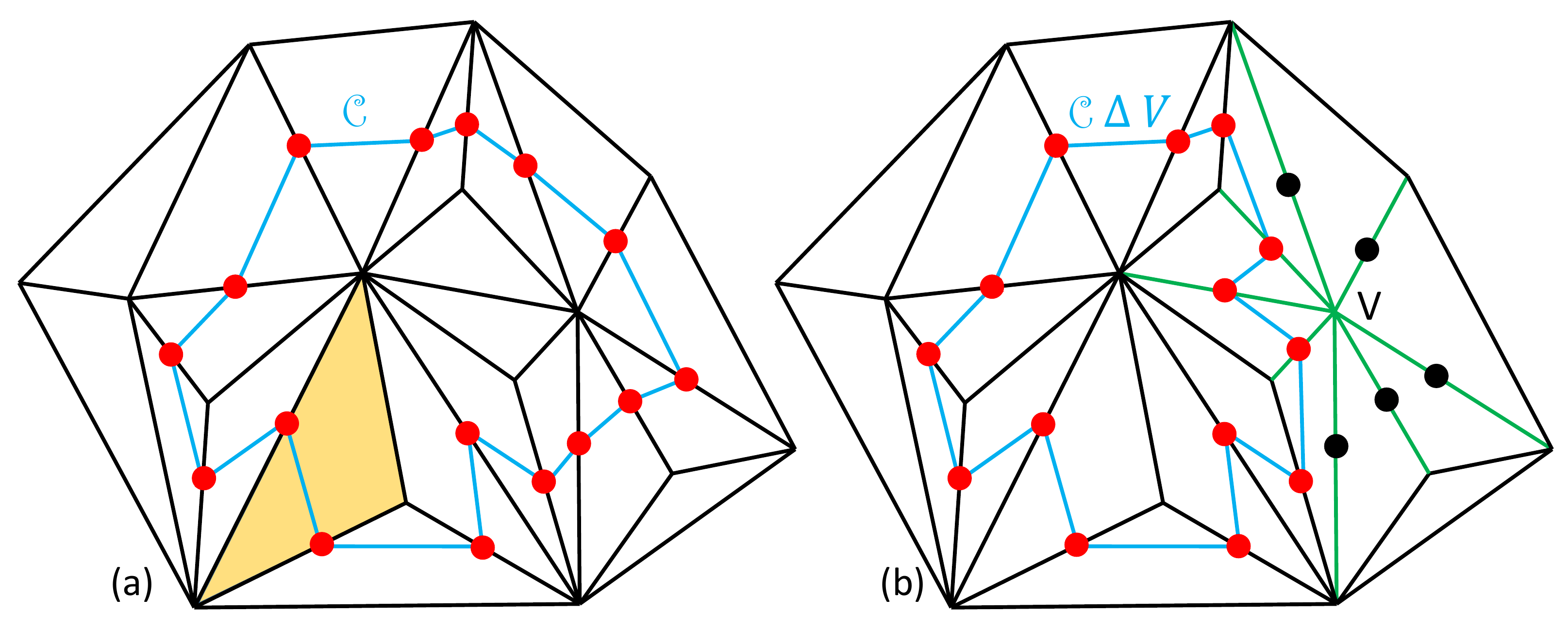}
\caption{\small (a) Example of an admissible contour $\Cc$, shown in blue. The contour skips from one fermion site to another in straight legs. The fermion sites are shown as red dots. Every single leg of the contour is contained in one and only one triangle, like in the one singled out with shading. (b) The symmetric difference of $\Cc$ with the vertex shown in green leads to the new contour $V \Delta \Cc$ shown again in blue, which is also an admissible contour. For clarity, the diagram shows as black dots the fermion sites which belonged to $\Cc$ and were dropped out by the finite difference.}
\label{Fig:Triangulation2}
\end{figure}

\begin{proposition}\label{Pro:Contour1} An allowed contour is a union of one or more non-intersecting, non-branching continuous lines. These lines are all necessarily closed.
\end{proposition}

\noindent {\bf Proof.} If a contour would display branching, then it would necessarily have to intersect all three edges of a triangle. This is prohibited by the rules. Self-intersection requires branching, hence it cannot occur, too. If a contour has a loose end, then that contour intersects a triangle on only one edge. This is also prohibited by the rules. $\square$

\vspace{0.2cm}

Contours which satisfy these rules are shown in Fig.~\ref{Fig:Triangulation2}. Note that all vertices $V\in \Ll$, more precisely the contours passing through the edges of the vertices, are part of the allowed contours. On surfaces of higher genus, the allowed contours can have non-trivial topological type. Such contours are shown in Fig.~\ref{Fig:Triangulation3} for the case of a torus. The aim of the following sections is to completely characterize the set $\mathfrak C$ of the allowed contours. 

\subsection{The loop sub-group $\GM_\ell$ defined}

\begin{proposition}\label{Pro:Stability} If $\Cc$ and $\Cc'$ are two admissible contours, then $\Cc \Delta \Cc'$ is also an admissible contour (or the empty set).
\end{proposition}

\noindent {\bf Proof.} Clearly, all legs of the symmetric difference are contained inside the triangles. Now, recall that the symmetric difference can be also written as:
\begin{equation}
\Cc \Delta \Cc' = (\Cc \cup \Cc') \setminus (\Cc\cap \Cc'),
\end{equation}
and that intersection operation distributes over $\Delta$. Then, for any triangle $T\in \Ll$:
\begin{equation}
T\cap (\Cc \Delta \Cc') = (T \cap \Cc) \Delta (T\cap \Cc'),
\end{equation}
and:
\begin{align}
|T\cap (\Cc \Delta \Cc')| & =|(T \cap \Cc) \Delta (T \cap \Cc')| \\ \nonumber
& = |(T \cap \Cc) \cup (T \cap \Cc') - (T \cap \Cc) \cap (T \cap \Cc')| \\ \nonumber 
& = |(T \cap \Cc)|+  |(T \cap \Cc')| - 2 |(T \cap \Cc) \cap (T \cap \Cc')|,
\end{align}
which shows that $\Cc \Delta \Cc'$ intersects $T$ at an even number of fermion sites. Since $|T|=3$, this number can only be zero or two. Hence, $\Cc \Delta \Cc'$ either does not intersect $T$ or it intersects precisely two edges of $T$. 

\vspace{0.2cm}

Here is a more wordy proof. If $\Cc$ and $\Cc'$ are non-intersecting, then we see from above that the statement is trivial. If they are intersecting, then $\Cc \cup \Cc'$ displays branching and let us focus on one of the branching points, where all three edges of a triangle are intersected by $\Cc \cup \Cc'$. One and only one of the tree intersection points belong to $\Cc \cap \Cc'$. This point is removed when taking the symmetric difference and two intersection points remain and the two contours can be joined in an allowed fashion. Hence, all the branch points are safely removed and a new allowed contour emerges. After this process, there might be additional points left in $\Cc \cap \Cc'$, such as when $\Cc$ and $\Cc'$ have common strings of legs, but all these points belong to $\Cc \cap \Cc'$ and are removed when taking the symmetric difference. $\square$

\begin{figure}
\center
\includegraphics[width=0.8\linewidth]{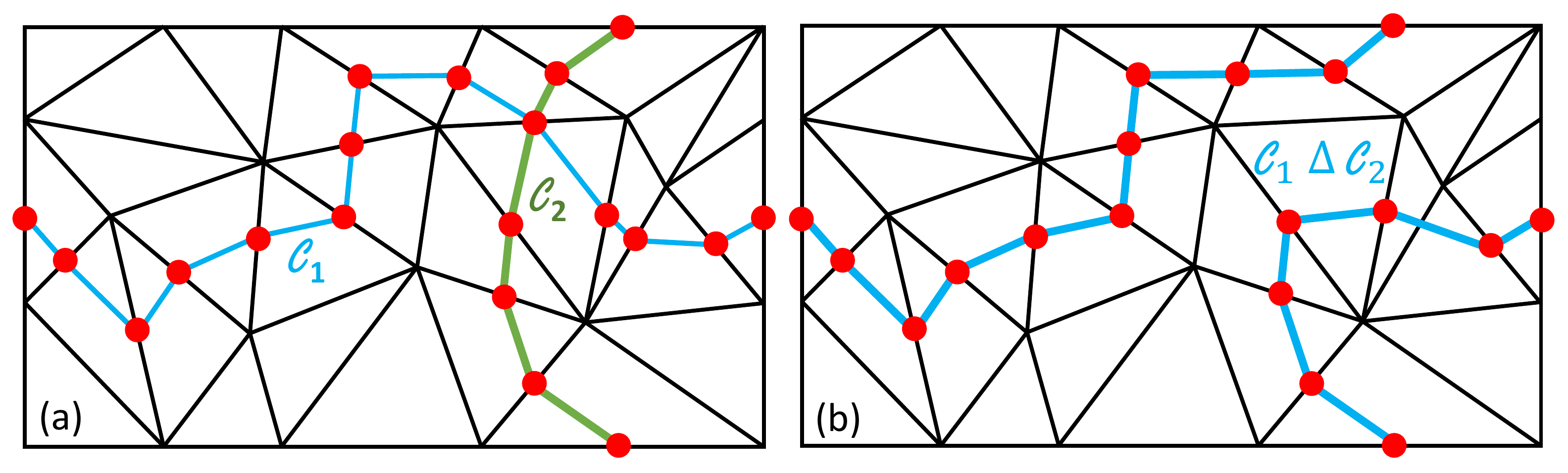}
\caption{\small (a) A triangulation of the (opened) torus and the two contours $\Cc_1$ and $\Cc_2$ from $\mathfrak C$ that cannot be reduced to the empty set by taking symmetric differences with vertices. (b) The symmetric difference between $\Cc_1$ and $\Cc_2$ results in an admissible contour, which is also topologically nontrivial.}
\label{Fig:Triangulation3}
\end{figure}

\vspace{0.2cm}

In Fig.~\ref{Fig:Triangulation2}, we show the symmetric difference between $\Cc$ and a vertex $V$. In this example, the contours share several consecutive legs and there are two branching points. In Fig.~\ref{Fig:Triangulation3}, we show the symmetric difference between $\Cc_1$ and $\Cc_2$. In this case, the intersection of the two contains a single fermion site, hence the branching is elementary. We suggest to the reader to focus on the branching points of the reunion and examine the mechanism that removes them when taking the symmetric difference. 

\begin{definition} Proposition~\ref{Pro:Stability} says that the set $\mathfrak C$ of admissible contours is closed under the  symmetric difference. This allows us to define the sub-group of loops:
\begin{equation} 
\GM_\ell = (\mathfrak C, \Delta) \subset (2^X,\Delta).
\end{equation}
We recall that the unit is just the empty set.
\end{definition}

\begin{remark}{\rm Given the isomorphism \eqref{Eq:MainIso}, $\GM_\ell$ defines a sub-group of $\bm S(X)$ as well, which will be denoted by the same symbol.
}$\Diamond$
\end{remark} 

\subsection{The lattice of sub-groups of $\GM_\ell$}

\begin{definition} A lattice is a partially ordered set in which every two elements have a unique supremum and a unique infimum.
\end{definition}

\begin{remark}{\rm The set of normal sub-groups of a group is partially ordered w.r.t. to set inclusion and, furthermore, can be organized in a lattice as it follows: The unique infimum of two normal sub-groups is their intersection and the unique supremum is their multiplication. 
}$\Diamond$
\end{remark}

Since $\GM_\ell$ is abelian, every sub-group is a normal sub-group. Since $\GM_\ell$ has order 2, every subset of the form $\{\emptyset,\Cc\}$, $\Cc \in \mathfrak C$, is a sub-group of $\GM_\ell$. In particular, $\{\emptyset, V\}$, $V \in \Ll$, are all sub-groups of $\GM_\ell$. We will build a useful sub-lattice starting from these sub-groups and more. First, a useful notation is needed. A finite group generated by a set $g_1, \ldots,g_n$ subjected to some relations $\Rr$ will be specified as:
\begin{equation}
\langle g_1,\ldots,g_n; \, \Rr \rangle.
\end{equation}
For example, $\{\emptyset, V\}$ becomes $\langle V \rangle$ in this notation. Note that the inclusion of the unit (which is $\emptyset$ here) is not specified explicitly by the notation. Furthermore, the product between two sub-groups results in the larger sub-group generated by the products of all elements of the sub-groups. For example:
\begin{equation}
\langle V \rangle \Delta \langle V' \rangle = \{\emptyset, \emptyset \Delta V', V \Delta \emptyset, V \Delta V'\} = \{\emptyset, V', V , V \Delta V'\}.
\end{equation}
As a side remark, one will be tempted to say that the right side is just $\langle V,V' \rangle$ but we will see below that this is not always the case since a relation might need to be specified. 

\begin{figure}
\center
\includegraphics[width=0.8\linewidth]{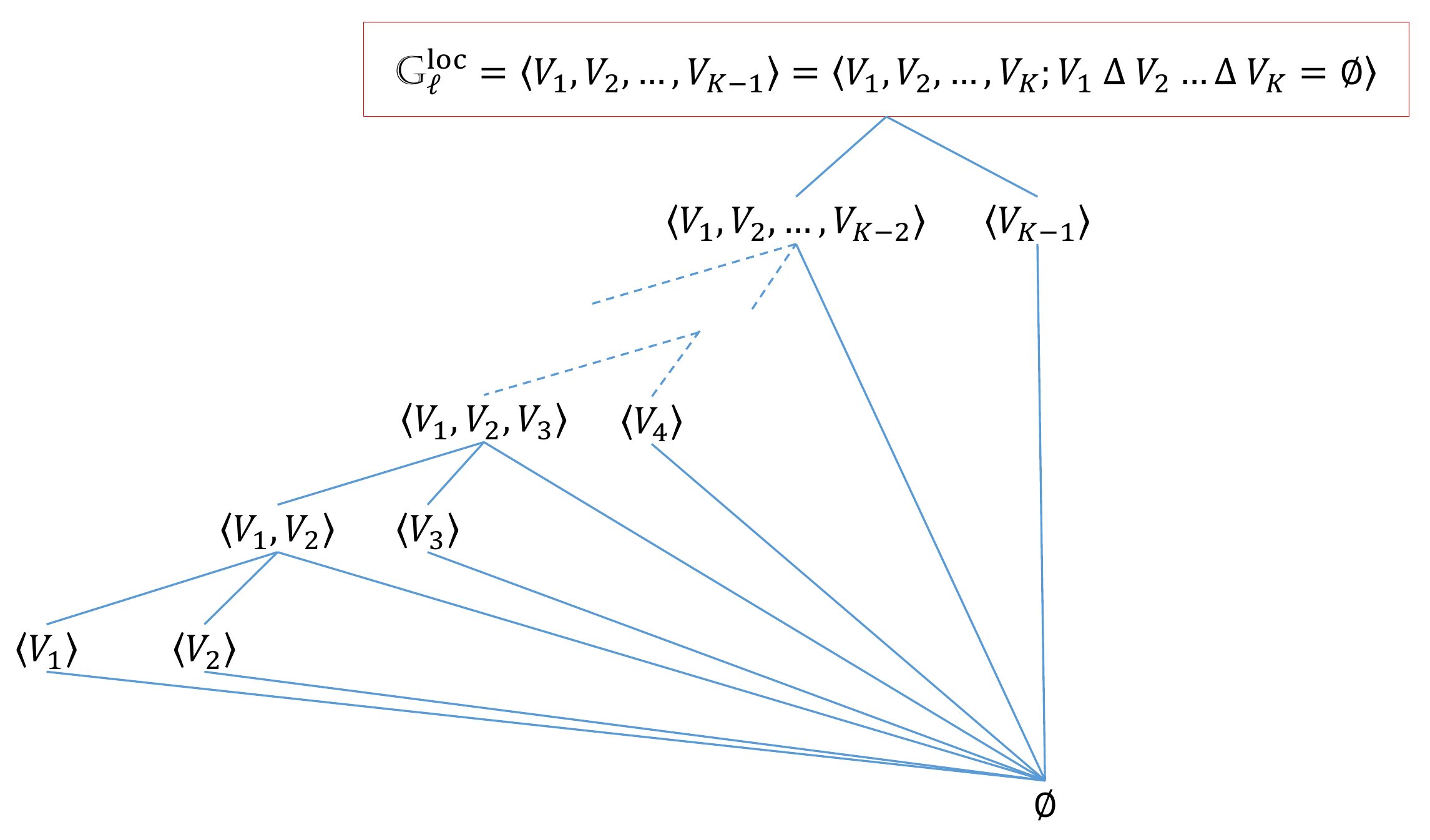}
\caption{\small A filtration of $\GM_\ell^{\rm loc}$ by a lattice of sub-groups. The unique sub-group above each pair of sub-groups is supplied by multiplication of the pair. The unique sub-group below each pair of sub-groups is supplied by the intersection of the pair. The sub-groups are ordered by inclusion as one goes up along any continuous path of the lattice.}
\label{Fig:Lattice1}
\end{figure}

\vspace{0.2cm}

We now define $\GM_\ell^{\rm loc}$ to be lowest subgroup above all $\{\emptyset, V\}$, $V \in \Ll$, that is:
\begin{equation}
\GM_\ell^{\rm loc} = \langle V \rangle \, \Delta \,  \langle V' \rangle \ldots,
\end{equation}
where the multiplication is over all vertices of the triangulation. This sub-group is of great importance because the algebra which supplies the Hamiltonians is generated by the elements of $\GM_\ell^{\rm loc}$:
\begin{equation}\label{Eq:NiceNot}
C^\ast(\Gamma_T,S_V; \, T,V \in \Ll) = C^\ast(\bm \Gamma_T,\GM_\ell^{\rm loc}).
\end{equation}
It will be helpful to label the vertices of $\Ll$ as $V_1$, \ldots, $V_K$.

\begin{proposition} In a triangulation:
\begin{equation}\label{Eq:Z1}
V_1 \Delta V_2 \ldots \Delta V_K = \emptyset.
\end{equation}
Furthermore, if the triangulated surface has only one connected component, then this is the only relation between the vertex loops and, as a consequence:
\begin{equation}
\GM_\ell^{\rm loc} = \langle V_1,\ldots,V_K; \,  V_1 \Delta V_2 \ldots \Delta V_K = \emptyset \rangle.
\end{equation}
\end{proposition}

\noindent {\bf Proof.} In a triangulation, every edge is shared between one and only one pair of vertices. Since in \eqref{Eq:Z1} the product is over all vertices, an edge appears in the product exactly two times and since $2^X$ is order 2 the first statement follows. If there are other combinations of symmetric differences that result in the empty set, then the vertices involved in such product cannot share any of their legs with vertices that are not involved in the product. This is because those legs will appear with a first power and cannot be reduced to the empty set. But if the vertices involved in the product do not share any of their legs with the remaining vertices, it means they triangulate a connected component of the surface. Since the only connected component is the surface itself, the remaining statements follow. $\square$

\begin{proposition} Regardless of the genus of the triangulated surface, the loops of $\GM_\ell^{\rm loc}$ are all contractible, hence of trivial topological type.
\end{proposition}

\noindent {\bf Proof.} We will use the filtration of $\GM_\ell^{\rm loc}$ by the lattice of subgroups shown in Fig.~\ref{Fig:Lattice1}. Definitely the vertex loops, in particular $V_1$, are contractible. By examining Fig.~\ref{Fig:Triangulation2}(b), we see that the effect of taking the symmetric difference with a vertex results in a local deformation of the contour. Hence, the loops of $\langle V_1,V_2 \rangle$ are all contractible and, by induction, the loops of $\langle V_1, \ldots V_{K-1} \rangle$ are all contractible. $\square$

\begin{figure}
\center
\includegraphics[width=0.4\linewidth]{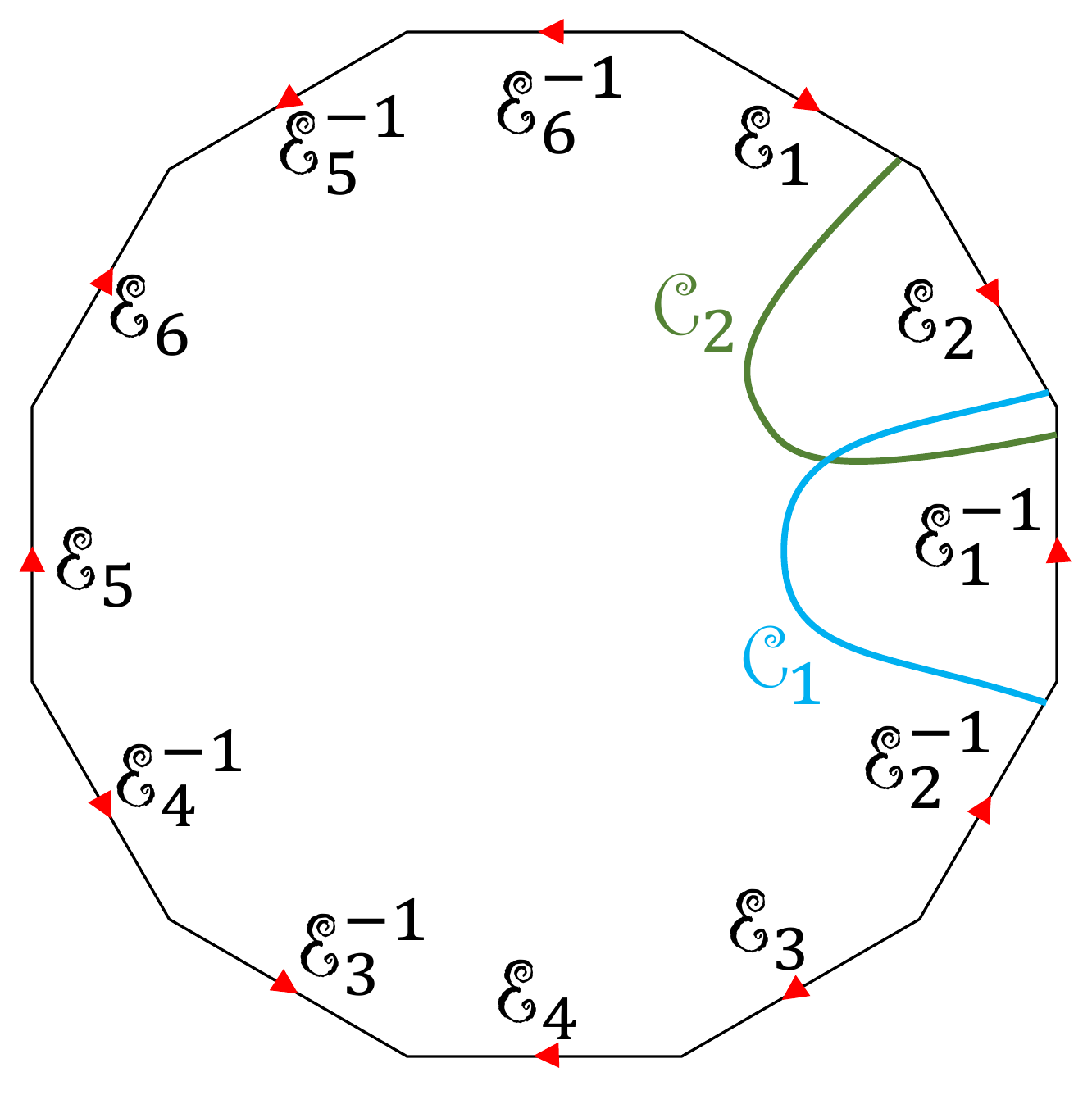}
\caption{\small A surface of genus $g$ ($g=3$ here) can be open to a $4g$-gon using $2g$ cuts originating from the same point. The corners of the $4g$-gon coincide with this point and are identified all together. Its edges supply a system of generating loops for surface's fundamental group. A pair of edges $(\Ee_i,\Ee_i^{-1})$ originate from the same cut. For reader's convenience, we show how the contours $\Cc_1^{-1}$ and $\Cc_2$ defined in Fig.~\ref{Fig:Triangulation3} will look on this diagram and that indeed they can be deformed into the $\Ee_1^{-1}$ and $\Ee_2$ edges.}
\label{Fig:GenusG}
\end{figure}

\vspace{0.2cm}

When the genus of the triangulated surface is zero, we have the equality $\GM_\ell = \GM_\ell^{\rm loc}$ but this is, of course, not the case for higher genus surfaces, where there are contours of non-trivial topological type, like $\Cc_1$ and $\Cc_2$ in Fig.~\ref{Fig:Triangulation3}(a), in the case of a torus. Let us recall that a general genus $g$ orientable surface $M$ can be opened to a $4g$-gon via $2g$ cuts that can start from the same point of $M$. As usual, we orderly label the edges of the $4g$-gon as:
\begin{equation}\label{Eq:Edge}
\Ee_1\Ee_2\Ee_1^{-1} \Ee_2^{-1} \ldots \Ee_{2g-1}\Ee_{2g}\Ee_{2g-1}^{-1}\Ee_{2g}^{-1},
\end{equation} 
where $\Ee_i$ and $\Ee_i^{-1}$ originate from the $i$-th cut, hence they have opposite orientations. The situation is illustrated in Fig.~\ref{Fig:GenusG}. Then the fundamental group $\pi_1(M)$ can be characterized as the free group generated by $\Ee_1$, \ldots, $\Ee_{2g}$ quoted by the normal group generated by the element spelled in \eqref{Eq:Edge}, or with our notation \cite[pp.~168]{ArmstrongBook}:
\begin{equation}
\pi_1(M) = \langle \Ee_1,\Ee_2, \ldots,  \Ee_{2g}; \Ee_1\Ee_2\Ee_1^{-1} \Ee_2^{-1} \ldots \Ee_{2g-1}\Ee_{2g}\Ee_{2g-1}^{-1}\Ee_{2g}^{-1}=1\rangle.
\end{equation}
The Abelianization of $\pi_1(M)$ then is:
\begin{equation}
\pi^a_1(M) = \langle \Ee_1,\Ee_2, \ldots,  \Ee_{2g}\rangle \simeq \ZM^{\times 2g},
\end{equation}
and the Booleanization is:
\begin{equation}
\pi^b_1(M) = \langle \Ee_1,\Ee_2, \ldots,  \Ee_{2g}; \Ee_1^2=\Ee_2^2=\ldots =\Ee_{2g}^2=1\rangle \simeq \ZM_2^{\times 2g}.
\end{equation}
Above, $\Ee_i$'s actually represent topological classes and each of these topological classes can be represented by an admissible contour from $\mathfrak C$, which we will denote by the symbol $\Cc_i$. The following statement now become obvious.

\begin{figure}
\center
\includegraphics[width=0.8\linewidth]{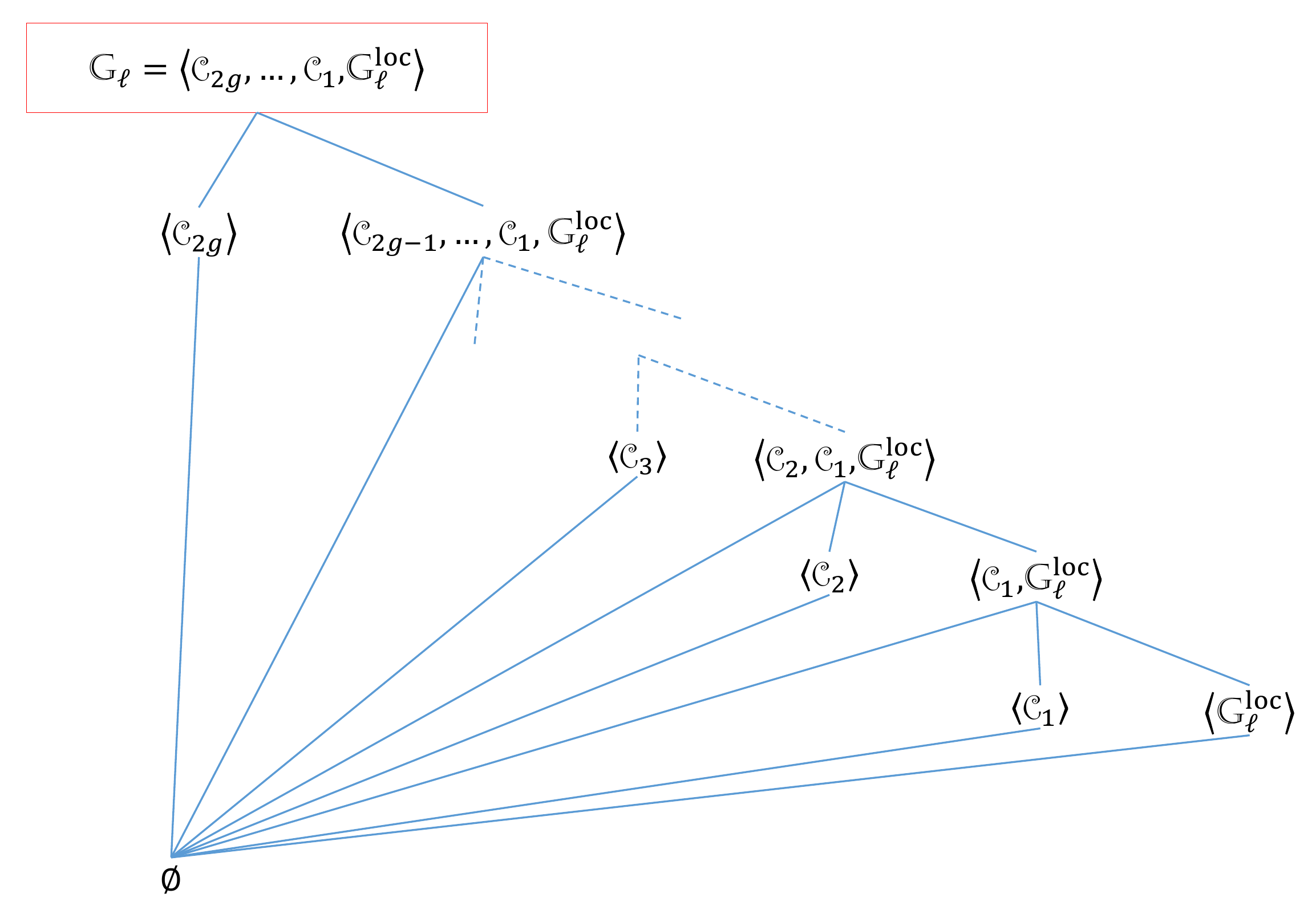}
\caption{\small A a lattice of sub-groups with $\GM_\ell$ as the upper bound and $\GM_\ell^{\rm loc}$ as the lower bound. All elements are of order 2.}
\label{Fig:Lattice2}
\end{figure}

\begin{proposition} For a triangulation of a surface of genus $g$, the group $\GM_\ell$ can be filtered through the lattice of subgroups shown in Fig.~\ref{Fig:Lattice2}. From this lattice, we can read the following property:
\begin{equation}\label{Eq:KeyProp}
\big ( \Cc_{j+1} \, \Delta \langle \Cc_j,\ldots,\Cc_1,\GM_\ell^{\rm loc} \rangle \big ) \cap \langle \Cc_j,\ldots,\Cc_1,\GM_\ell^{\rm loc} \rangle = \emptyset, \quad \forall \ j=0,\ldots, 2g-1.
\end{equation}
Furthermore:
\begin{equation}
\GM_\ell / \GM_\ell^{\rm loc} \simeq \pi_1^b(M).
\end{equation}
\end{proposition}

Using the isomorphism \eqref{Eq:MainIso}, we can transform the lattice in Fig.~\ref{Fig:Lattice2} into a lattice of sub-groups of $\bm S(X)$. More specifically, we define the symmetries:
\begin{equation}
\Sigma^i = S_{\Cc_i} = \prod_{x \in \Cc_i} s_x \in CAR(X), \quad i=1,\ldots, 2g,
\end{equation}
in which case we have the filtration:
\begin{equation}\label{Eq:SFiltration}
\GM_\ell^{\rm loc} \subsetneq \langle \Sigma^1,\GM_\ell^{\rm loc} \rangle \ldots \subsetneq \langle \Sigma^{2g},\ldots, \Sigma^1,\GM_\ell^{\rm loc} \rangle = \GM_\ell,
\end{equation}
with the obvious but crucial property:
\begin{equation}\label{Eq:GCrucial}
\big (\Sigma^{j+1} \cdot \langle \Sigma^j,\ldots, \Sigma^1,\GM_\ell^{\rm loc} \rangle \big ) \cap \langle \Sigma^j,\ldots, \Sigma^1,\GM_\ell^{\rm loc} \rangle = \emptyset.
\end{equation}

\begin{remark}{\rm Another way to look at these properties is to realize that on the very last path of the lattice we have a directed tower of normal sub-groups. Then the lattice simply keeps track of the long exact group sequences:
\begin{equation}
\emptyset \rightarrow \langle \Sigma^{j+1} \rangle \rightarrow \langle \Sigma^{j+1},\ldots, \Sigma^1,\GM_\ell^{\rm loc} \rangle \rightarrow \langle \Sigma^{j},\ldots, \Sigma^1,\GM_\ell^{\rm loc} \rangle \rightarrow \emptyset.
\end{equation}
It appears to us that our arguments leading to the topological spectral degeneracy work for generic towers of exact sequences. As such, we hope that topological spectral  degeneracy will be searched and found for many other lattices of sub-groups of $2^X$.
}$\Diamond$
\end{remark}

\section{Topological Models, Spectral Degeneracy}

In this section we prove points i) and ii) of Theorem~\ref{Th:Main}. Henceforth, we consider the proper sub-algebra of $CAR(X)$ generated by the elementary elements of $\Ll$, namely:
\begin{equation}\label{Eq:GoodRep}
C^\ast(\Gamma_T, S_V; \, T,V \in \Ll) = C^\ast(\GM_\ell^{\rm loc},\bm \Gamma_T).
\end{equation}
We will show that any Hamiltonian drawn from this sub-algebra displays topological spectral degeneracy. More precisely, let $H$ be a self-adjoint element of the CAR-algebra generated by arbitrary linear combinations of products of $\Gamma_T$'s and $S_V$'s, which we convey with the notation:
\begin{equation}
H = H(\Gamma_T,S_V; \, T,V\in \Ll).
\end{equation}
Let $\Pp$ be the spectral projection corresponding to an eigenvalue of $H$. Quite straightforward arguments will show that, for any triangulation of a higher genus surface, $\Pp$ cannot be a minimal projection and that $\Pp$ decomposes in lower projection as stated at point i) of Theorem~\ref{Th:Main}. Key to the proof is an embedding of the Booleanization of $\pi_1(M)$ inside the group of unitary elements of the corner sub-algebra $\Pp CAR(X) \Pp$.

\begin{remark}{\rm We recall, {\it e.g.} from \cite[Sec.~1.6]{JonesBook1}, that two projection $\Pp$ and $\Qq$ are ordered as $\Pp \leq \Qq$ iff there exists a partial isometry $U$ from the same algebra such that $UU^\ast = \Pp$ and $U^\ast U \Qq = U^\ast U$. Furthermore, if the algebra is a factor as is the case for $CAR(X)$, then any two projections can be compared. A projection is called minimal if it dominates no other projection except $0$.
}$\Diamond$
\end{remark}

\subsection{The commutant algebra argument}

As stated above, we want to investigate the structure of the spectral projection $\Pp$, more precisely, if there is any projection $P$ in $CAR(X)$ such that $P \Pp = \Pp P = P$ and $P \neq \Pp$. If the answer is yes, then $\Pp$ is not minimal, hence it can be decomposed in direct sum of smaller projections, namely $\Pp = P\Pp +(1-P)\Pp$. This implies degeneracy of the eigenvalue. Of course, to get to the full statement in Theorem~\ref{Th:Main}, the argument needs to be iterated.

\vspace{0.2cm}

 In \cite{KitaevAOP2003}, Kitaev advises that one should look at the commutant algebra of the sub-algebra $C^\ast(\Gamma_T, S_V; \, T,V \in \Ll)$ inside the algebra of observables, in our case, $CAR(X)$. This commutant algebra is a sub-algebra of $CAR(X)$, defined as:\footnote{Technically, this is the relative commutant.}
\begin{align}
C^\ast(\Gamma_T, S_V; \, T,V \in \Ll)' = &  \{A \in CAR(X)\ | \\ \nonumber 
&  \quad A B = B A, \ \forall \, B \in C^\ast(\Gamma_T, S_V; \, T,V \in \Ll)\}.
\end{align}
As opposed to the toric code, the commutant algebra no longer includes the sub-algebra $C^\ast(\Gamma_T, S_V; \, T,V \in \Ll)$. Nevertheless, the standard procedure leading to the topological degeneracy will be to identify a sub-algebra of the commutant, which accepts only higher dimensional irreducible representations.

\vspace{0.2cm}

We want, however, to present a new strategy. We will search for a symmetry $\Sigma$, {\it i.e.} $\Sigma^\ast = \Sigma$, $\Sigma^2=1$, from the commutant $C^\ast(\Gamma_T, S_V; \, T,V \in \Ll)'$, such that $\Sigma A$ does not belong to $C^\ast(\Gamma_T, S_V; \, T,V \in \Ll)$ for any $A \in C^\ast(\Gamma_T, S_V; \, T,V \in \Ll)$, $A\neq 0$. In short:
\begin{equation}\label{Eq:Arg2}
\Big ( \Sigma \cdot C^\ast(\Gamma_T, S_V; \, T,V \in \Ll) \Big ) \cap C^\ast(\Gamma_T, S_V; \, T,V \in \Ll) = \{0\}.
\end{equation}
The spectral projections $\Sigma_\pm$ of $\Sigma$, defined by $\Sigma=\Sigma_+ - \Sigma _-$, $1=\Sigma_+ + \Sigma_- $, commute with $\Pp$ and will supply the decomposition of $\Pp$ in more elementary projections: 
\begin{equation}\label{Eq:Arg1}
\Pp = 1 \cdot \Pp = (\Sigma_+ + \Sigma_-)\Pp =  \Sigma_+ \Pp + \Sigma_- \Pp,
\end{equation} 
and, obviously:
\begin{equation}\label{Eq:Arg11}
(\Sigma_\pm \Pp)^2= \Sigma_\pm \Pp, \quad (\Sigma_+ \Pp)(\Sigma_- \Pp)=0.
\end{equation} 
For this to be meaningful, we need to make sure that $\Sigma_\pm \Pp \neq 0$, and this can be derived as follows. Being a spectral projection of $H$, $\Pp$ belongs to $C^\ast(\Gamma_T, S_V; \, T,V \in \Ll)$ and \eqref{Eq:Arg2} says that necessarily $\Sigma \Pp \neq \pm \Pp$. But if one of $\Sigma_\pm \Pp$ are zero, say $\Sigma_- \Pp=0$, then: 
\begin{equation}
\Pp = 1\cdot \Pp = (\Sigma_+ + \Sigma_-)\Pp= \Sigma_+ \Pp= (\Sigma_+  - \Sigma_-)\Pp  = \Sigma \Pp.
\end{equation}
The contradiction proves that the assumption $\Sigma_-\Pp=0$ cannot be true.

\vspace{0.2cm}

Since this argument will be iterated, it is useful to formulate the core idea in a context-free statement:

\begin{proposition}\label{Pro:CoreArg} Let $\Bb$ be a $C^\ast$-algebra, $\Aa \subsetneq \Bb$ a proper sub-algebra and $\Aa'$ the commutant algebra of $\Aa$ inside $\Bb$. Let $\Sigma\in \Bb$ such that:
\begin{itemize}
\item $\Sigma$ is invertible and self-adjoint;
\item $\Sigma$ is an element of the commutant $\Aa'$;
\item $\Sigma \cdot \Aa \cap \Aa = \{0\}$. 
\end{itemize}
Then, if $\Sigma_\pm$ represent the spectral projections onto the positive/negative parts of the spectrum of $\Sigma$, then, for any projection $P$ from $\Aa$:
\begin{equation}
P = \Sigma_+P + \Sigma_- P,
\end{equation}
where $\Sigma_\pm P = P \Sigma_\pm$ are mutually orthogonal projections:
\begin{equation}
(\Sigma_\pm P)^2 = (\Sigma_\pm P), \quad (\Sigma_+P)(\Sigma_-P)=0,
\end{equation}
and each of them must be nonzero, $\Sigma_\pm P \neq 0$.
\end{proposition} 
 
 \subsection{Topological spectral degeneracy demonstrated}
 
 We will switch to the more compact notation of \eqref{Eq:GoodRep}. Since the admissible contours from $\mathfrak C$ intersect either zero or two edges of triangle, it follows from \eqref{Eq:CommGS} that any element $S_{\Cc}$ from $\GM_\ell$ commutes with the elementary triangle elements. In other words, all $S_\Cc$ belong to commutant $C^\ast(\GM_\ell^{\rm loc},\bm \Gamma_T)'$.
 
 \begin{proposition} There is the filtration of $C^\ast$-algebras:
 \begin{align}\label{Eq:CoreFilter}
& C^\ast(\Gamma_T, S_V; \, T,V \in \Ll) = C^\ast(\GM_\ell^{\rm loc},\bm \Gamma_T) \subsetneq C^\ast(\Sigma^1,\GM_\ell^{\rm loc},\bm \Gamma_T) \\ \nonumber
& \qquad \qquad \subsetneq C^\ast(\Sigma^{2},\Sigma^{1},\GM_\ell^{\rm loc},\bm \Gamma_T) \ldots \subsetneq C^\ast(\Sigma^{2g},\ldots,\Sigma^{1},\GM_\ell^{\rm loc},\bm \Gamma_T) ,
 \end{align}
 with the additional properties:
 \begin{itemize}
 
 \vspace{0.1cm}
 
 \item $\Sigma^{j+1} \in C^\ast(\Sigma^{j},\ldots,\Sigma^{1},\GM_\ell^{\rm loc},\bm \Gamma_T)' ;$
 
 \vspace{0.1cm}
 
 \item $ \big ( \Sigma^{j+1} \cdot C^\ast(\Sigma^j,\ldots,\Sigma^1,\GM_\ell^{\rm loc},\bm \Gamma_T) \big ) \cap C^\ast(\Sigma^j,\ldots,\Sigma^1,\GM_\ell^{\rm loc},\bm \Gamma_T) = \{0\},$
 \end{itemize}
 for all $j=0,\ldots 2g-1$.
 \end{proposition}
 
\noindent {\bf Proof.} The filtration written in \eqref{Eq:CoreFilter} can be translated in the language of crossed product algebras:
 \begin{align}
\bm \Gamma_T \rtimes \GM_\ell^{\rm loc}  \subsetneq \bm \Gamma_T \rtimes \langle \Sigma^1,\GM_\ell^{\rm loc}\rangle \ldots  \subsetneq \bm \Gamma_T \rtimes \langle \Sigma^{2g},\ldots,\Sigma^{1},\GM_\ell^{\rm loc} \rangle,
 \end{align} 
and \eqref{Eq:CoreFilter} follows directly from \eqref{Eq:SFiltration}. The second statement was already covered above and the third one follows directly from Remark~\ref{Re:ZeroA} and \eqref{Eq:GCrucial}. $\square$

\vspace{0.2cm}
 
We are now ready to prove points i) and ii) of Theorem~\ref{Th:Main}. Henceforth, let $H$ be a self-adjoint element from $C^\ast(\GM_\ell^{\rm loc},\bm \Gamma_T)$ and $\Pp$ one of its spectral projections, which belongs to the same sub-algebra. By applying Proposition~\ref{Pro:CoreArg} on the first leg of the filtration \eqref{Eq:CoreFilter}, we can safely conclude that:
\begin{equation}
\Pp = \Sigma^1_+ \Pp + \Sigma^1_- \Pp.
\end{equation}
Now, the proper projections $\Sigma^1_\pm \Pp$ belong to $C^\ast(\Sigma^1,\GM_\ell^{\rm loc},\bm \Gamma_T)$, hence we can apply Proposition~\ref{Pro:CoreArg} on the second leg of the filtration \eqref{Eq:CoreFilter}. This leads to another decomposition in terms of proper projections:
\begin{equation}
\Sigma^1_\pm \Pp = \Sigma^2_+\Sigma^1_\pm \Pp + \Sigma^2_- \Sigma^1_\pm \Pp.
\end{equation}
The argument can be iterated for each leg of the filtration \eqref{Eq:CoreFilter} and the final conclusion is that $\Pp$ decomposes in $2^{2g}$ mutually orthogonal and non-zero projections:
\begin{equation}
\Pp = \sum_{\alpha_1=\pm} \ldots \sum_{\alpha_{2g}=\pm} \ \prod_{j=1}^{2g}\Sigma^j_{\alpha_j} \Pp.
\end{equation} 
This concludes point i). 

\vspace{0.2cm}

As for point ii), the embedding of $\pi_1^b(M)\simeq \ZM_2^{\times 2g}$ is supplied by the sub-group $\langle \Sigma^1 \Pp,\ldots,\Sigma^{2g}\Pp\rangle$ of the invertible elements of the corner sub-algebra $\Pp CAR(X) \Pp$. There is clearly a morphism between the two groups, which is injective because we just learned that all $\Sigma^j_\pm \Pp$'s are non-zero, hence none of the $\Sigma^i \Pp$ can be unity. 

\section{Structure of the algebra of the protected physical observables}

The algebra of protected physical observables is the corner sub-algebra $\Pp CAR(X) \Pp$. This algebra is also a linear space, which can be regarded as a computational space, in the context of topological quantum computation. In fact, the GNS representation w.r.t. state $\omega_\Pp$ (see \eqref{Eq:OmegaState}) transforms this linear space into a Hilbert space. We learn from points i) and ii) of Theorem~\ref{Th:Main} that this space is at least $4g$-dimensional and that it can be divided in $4g$ sub-spaces using the commuting observables $\Sigma^i\Pp$. In this section, we are searching for additional observables that act between these sub-spaces. More precisely, using the simple rule from Remark~\ref{Re:CP} for computing the trace, one can see that $\Tt(\Sigma^i)=0$ for all $i$'s, hence the spectral projections $\Sigma_\pm^i$ have the same dimension. This assures us that the projections $\Sigma^i_\pm$ can be intertwined and, as such, there is a unitary $\Xi^i \in CAR(X)$ such that $\Xi^i \Sigma^i \Xi^i = -\Sigma^i$. However, the crucial question is: Can we construct $\Xi$'s explicitly and inside the corner algebra $\Pp CAR(X) \Pp$? The statement iii) of Theorem~\ref{Th:Main} says that we can and the proof is supplied in this section.

\subsection{Admissible $\Gamma$-contours defined}

We will define a new set of contour-like elements of $CAR(X)$ that are associated with the $\gamma$-elements this time. These are supported by special contours $\Qq\subset X$ defined as:

\begin{definition}{\rm The rules for $\Gamma$-admissible contours are:
\begin{itemize}
\item A contour skips from one fermion site to another in a linear fashion.

\item Each leg of a contour coincides with an edge of $\Ll$.

\item The contour, as it meanders over the triangulation, shares an even number of edges with any of the vertices in $\Ll$.

\end{itemize} 

}
\end{definition}

\begin{proposition} An admissible $\Gamma$-contour can be decomposed into reunion of closed loops which can intersect only at vertices.
\end{proposition}

\noindent {\bf Proof.} A $\Gamma$-admissible contour cannot have a loose end because there will be a vertex sharing only one edge with the contour. We can then identify at least one closed loop and, once we did that, we remove it from the contour. This will remove either none or two legs from each of the vertices. Hence, the remainder is again a $\Gamma$-admissible contour and we can identify another closed loop. Being in the remainder, this new closed loop does not contained any edges of the first loop. The process can be iterated until the remainder is the empty set. $\square$

\begin{figure}
\center
\includegraphics[width=0.8\linewidth]{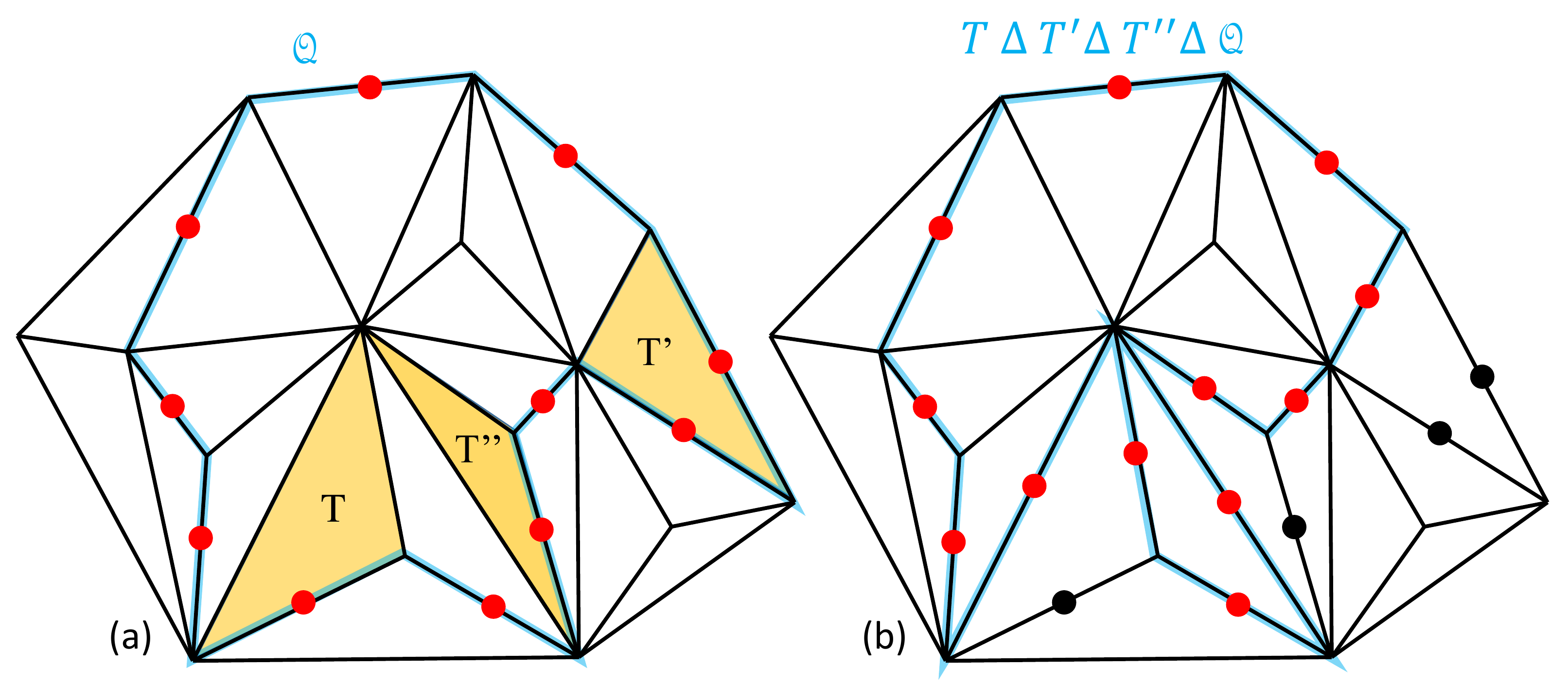}
\caption{\small (a) Example of a $\Gamma$-admissible contour, shown in blue. The contour skips from one fermion site to another in straight legs. The fermion sites are shown as red dots. Every single leg of the contour coincides with an edge of the triangulation. (b) The symmetric difference of $\Qq$ with the triangles shaded in panel (a) leads to a new contour, shown again in blue, which is again a $\Gamma$-admissible. For clarity, the diagram shows as black dots the fermion sites which belonged to $\Qq$ and were dropped out by the symmetric difference.}
\label{Fig:Triangulation4}
\end{figure}

\begin{proposition} If $\Qq$ and $\Qq'$ are $\Gamma$-admissible contours, then so is $\Qq \Delta \Qq'$. As such, the set $\mathfrak Q$ of all $\Gamma$-admissible contours defines a sub-group $(\mathfrak Q, \Delta) \subset (2^X,\Delta)$.
\end{proposition}

\noindent {\bf Proof.} For any vertex $V\in \Ll$:
\begin{equation}
V\cap (\Qq \Delta \Qq') = (V \cap Q) \Delta (V\cap \Qq'),
\end{equation}
hence:
\begin{align}
|V\cap (\Qq \Delta \Qq')| & =|(V \cap \Qq) \Delta (V \cap \Qq')| \\ \nonumber
& = |(V \cap \Qq) \cup (V \cap \Qq') - (V \cap \Qq) \cap (V \cap \Qq')| \\ \nonumber 
& = |(V \cap \Qq)|+  |(V \cap \Qq')| - 2 |(V \cap \Qq) \cap (V \cap \Qq')|,
\end{align}
which shows that $\Qq \Delta \Qq'$ share with $V$ an even number of edges. $\square$

\vspace{0.2cm} 

The elementary triangles belong to $\mathfrak Q$. Additional contours satisfying the above rules are shown in Fig.~\ref{Fig:Triangulation4} for a genus zero surface and in Fig.~\ref{Fig:Triangulation5}(a) for a genus 1 surface. The effect of symmetric difference is exemplified in Fig.~\ref{Fig:Triangulation4}, for triangles sharing one or two edges with a  contour $\Qq$. Since all the elementary triangles belong to $\mathfrak Q$, we can generate many contours from $\mathfrak Q$ by taking sequential symmetric differences of triangles and, on surfaces of genus $0$, this procedure generates all $\mathfrak Q$ but this is not the case on surfaces of higher genus. In Fig.~\ref{Fig:Triangulation5}(a), we illustrate, for the case of a torus, two special contours, $\Qq_1$ and $\Qq_2$, which cannot be reduced to the empty set by taking symmetric differences with triangles. For a generic surface of genus $g$, we can define $2g$ such contours that cannot be deformed one into another by symmetric differences with triangles. This is illustrated in Fig.~\ref{Fig:GenusGG}.

\begin{remark}{\rm As one can see, there is a duality between the $\mathfrak C$ and $\mathfrak Q$ contours but it is not a complete duality. The contours $\mathfrak C$ are closed non-intersecting loops, while $\mathfrak Q$ contours self-intersect, in general. As shown in Fig.~\ref{Fig:Triangulation4}(b), even if we start with non-intersecting contour $\Qq$, after taking the symmetric difference with triangles, the contour develops self-intersections at vertices. This was the reason we defined the $S$ elements over the $\mathfrak C$ and not $\mathfrak Q$ contours.
}$\Diamond$
\end{remark}

\subsection{A new set of elementary operators}

We define a new set of elementary operators from $CAR(X)$:
\begin{equation}
\xi_x = \imath \gamma_x s_x, \quad \xi_x^\ast = \xi_x, \quad \xi_x^2 = 1, \quad x \in X,
\end{equation}
satisfying the commutation relations:
\begin{align}
\xi_x \xi_{x'} & = \imath^2 \gamma_x s_x \gamma_{x'} s_{x'} \\ \nonumber
& = - \imath^2 (-1)^{\delta_{xx'}}\gamma_{x'} s_{x'}  \gamma_x s_x = -(-1)^{\delta_{xx'}} \gamma_{x'}\gamma_{x}, \quad \forall \, x,x'\in X,
\end{align}
and:
\begin{align}
\gamma_{x'} \xi_x \gamma_{x'} & = \imath  \gamma_{x'} \gamma_x s_x \gamma_{x'}=\imath (\gamma_{x'} \gamma_x \gamma_{x'}) \, (\gamma_{x'} s_x \gamma_{x'}) \\ \nonumber
& = -\imath (-1)^{ \delta_{xx'}} \gamma_x \, (-1)^{\delta_{xx'}} s_x =-  \xi_x,   \quad \forall \, x,x' \in X,
\end{align}
as well as:
\begin{align}
s_{x'} \xi_x s_{x'} & = \imath s_{x'} \gamma_x s_x s_{x'}=\imath s_{x'} \gamma_x s_{x'} s_x \\ \nonumber
&  =\imath (-1)^{\delta_{xx'}} \gamma_x s_x = (-1)^{\delta_{xx'}} \xi_x,   \quad \forall \, x,x' \in X.
\end{align}

\begin{remark}{\rm As one can notice, the commutation relations for $\xi$'s, among themselves and with the $s$ elements, are identical with those for $\gamma$ elements, which are listed in section \ref{Sec:ElementaryGen}.
}$\Diamond$
\end{remark}

\begin{proposition} The map:
\begin{equation}
2^X \ni J \mapsto \Xi_{J} = \imath^{\eta_{JJ}+|J|} \prod_{x \in J} \xi_x  \in CAR(X),
\end{equation}
defines a projective representation of $(2^X,\Delta)$ inside the group of invertible elements of $CAR(X)$.
\end{proposition}

\noindent {\bf Proof.} One can directly verify that $\Xi_J$'s square to the identity and, given that $\xi_x^2=1$, we have $\Xi_J \Xi_{J'} =(-1)^{\eta'_{J J'}} \Xi_{J \Delta J'} $, with $\eta'_{JJ'}$ a two-cocycle. $\square$

\vspace{0.2cm}

The following commutation relations will be essential for our final arguments:
\begin{equation}\label{Eq:Xi1}
\Xi_J \Xi_{J'} = (-1)^{|J| \cdot |J'|+|J\cap J'|} \Xi_{J'} \Xi_J, \quad \forall \, J,J' \in 2^X,
\end{equation}
and:
\begin{equation}\label{Eq:Xi2}
\Xi_J S_{J'} = (-1)^{|J \cap J'|} S_{J'} \Xi_J, \quad \forall \, J,J' \in 2^X,
\end{equation}
as well as:
\begin{equation}\label{Eq:Xi3}
\Xi_J \Gamma_{J'} = (-1)^{|J| \cdot |J'|} \Gamma_{J'} \Xi_J, \quad \forall \, J,J' \in 2^X.
\end{equation}
Hence, the commutation relations between $\Xi_J$'s and between $\Xi_J$'s and $S_J$'s are the same as the ones between $\Gamma_J$'s and between $\Gamma_J$'s and $S_J$'s written in \eqref{Eq:Gamma0} and \eqref{Eq:CommGS}, respectively. But there is an important difference when we compare the commutation relations between $\Xi_J$'s and $\Gamma_J$'s and the ones between the $\Gamma_J$'s, written in \eqref{Eq:Gamma0}, which will be exploit next.

\subsection{The structure revealed}

\begin{proposition} All elements $\Xi_\Qq$ with $\Qq \in \mathfrak Q$ and $|\Qq|$ even belong to the commutant of $C^\ast(\Gamma_T,S_V; \, T,S \in \Ll)$.
\end{proposition}

\noindent {\bf Proof.} From \eqref{Eq:Xi1}, it follows that any $\Xi_{\Qq}$ with $|\Qq|$ even commutes with any $\Gamma_J$, in particular, with the triangle elements. Furthermore, since $\Qq$'s share an even number of edges with any vertex, it follows from \eqref{Eq:Xi2} that $\Xi_{\Qq}$'s also commute with the vertex elements. $\square$ 

\begin{figure}
\center
\includegraphics[width=0.8\linewidth]{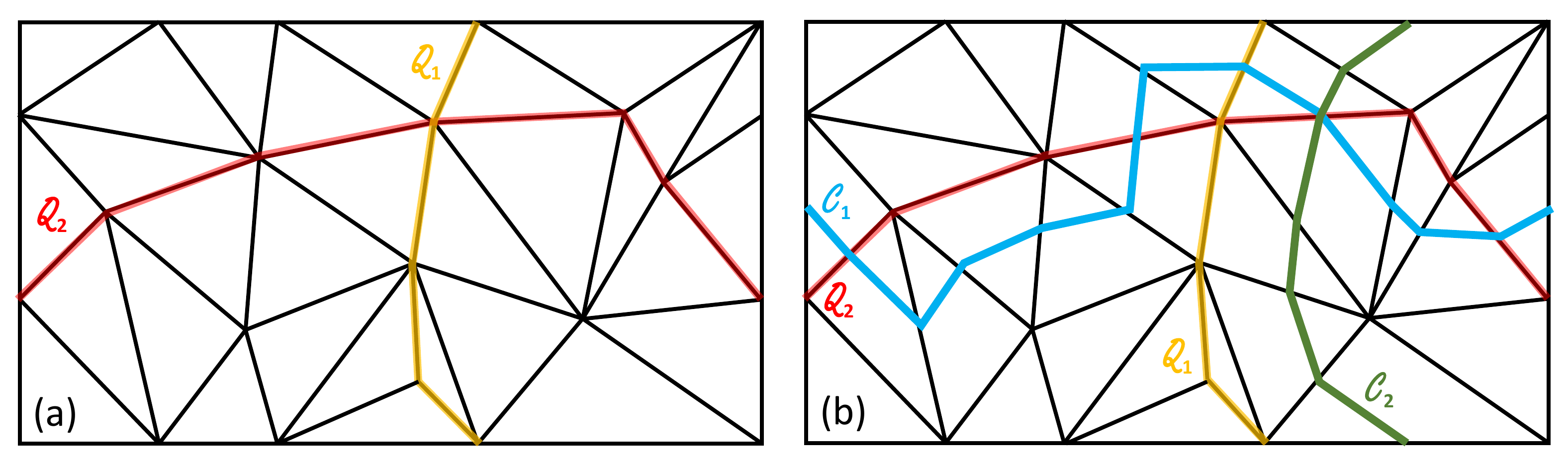}
\caption{\small (a) Examples of two $\Gamma$-admissible contours in a triangulation of a torus, which cannot be reduced to the empty set by taking symmetric differences with triangles. (b) Superposition of the topological $\mathfrak C$ and $\mathfrak Q$ contours.}
\label{Fig:Triangulation5}
\end{figure}

\vspace{0.2cm}

\begin{figure}
\center
\includegraphics[width=0.5\linewidth]{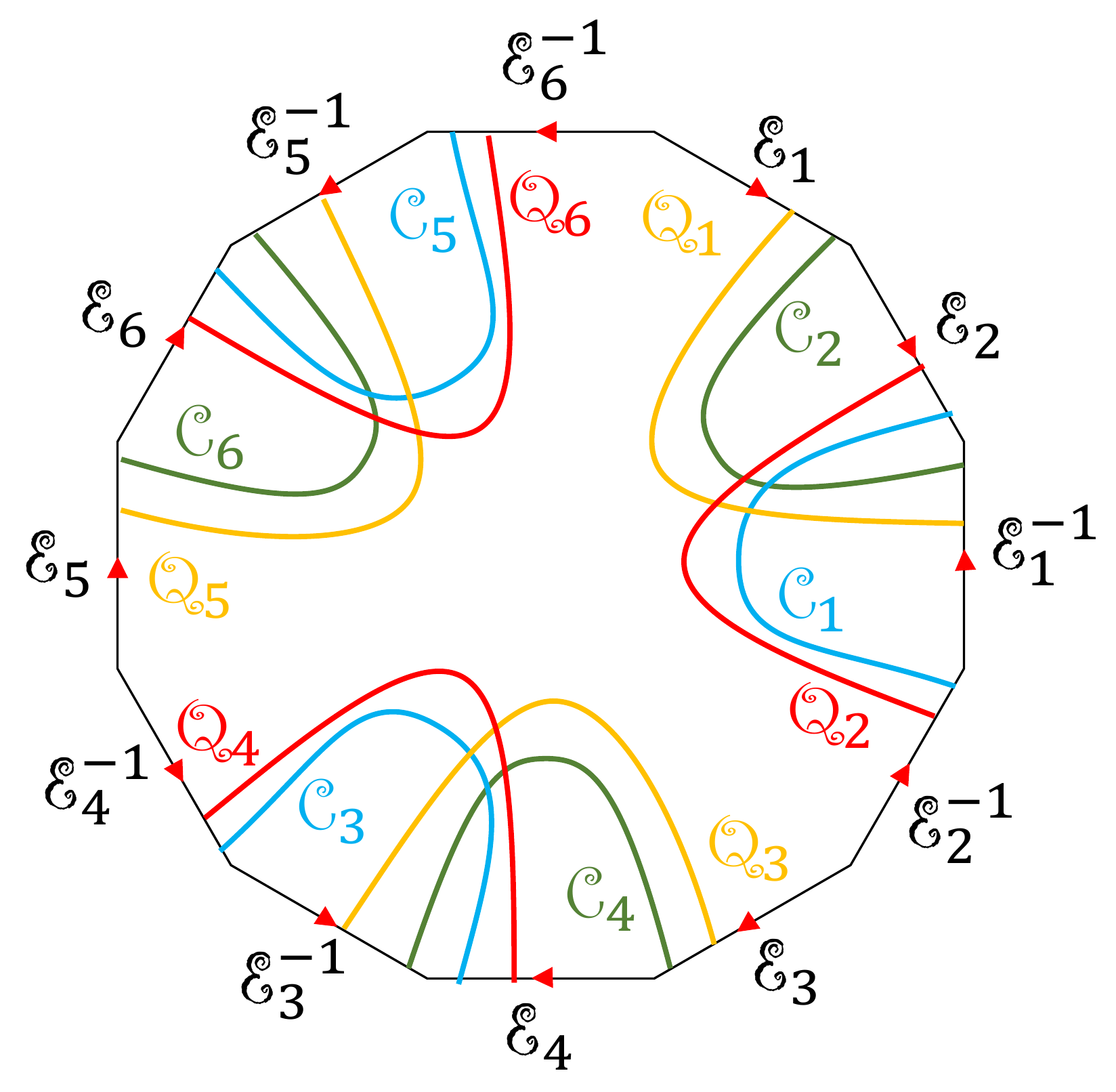}
\caption{\small Convenient choice of $\mathfrak C$ and $\mathfrak Q$ topological contours on a genus $3$ surface.}
\label{Fig:GenusGG}
\end{figure}

We now focus on the topological contours shown in Fig.~\ref{Fig:Triangulation5} for the torus and in Fig.~\ref{Fig:GenusGG} for a generic genus surface.  Examining the paths $\Qq_1$ and $\Qq_2$ shown in Fig.~\ref{Fig:Triangulation5}(a), one should observe the following special characteristics, which can be always put in place by refining the triangulation \footnote{By refining a triangulation, we mean placing one point inside one of the triangles and breaking that triangle into three smaller triangles that share a vertex at that point.}, if necessary:
\begin{itemize}

\item They have an even number of legs: $|\Qq_i|$ = even. 

\item The contours intersect transversely, {\it i.e.} only at one point. Since this point is necessarily a vertex and the fermions leave on the edges, $\Qq_1 \cap \Qq_2 = \emptyset$.

\end{itemize}
By examining Fig.~\ref{Fig:GenusGG}, it is quite clear that these characteristics can be also enforced on higher genus triangulations. With these choices, it then follows directly from \eqref{Eq:Xi1} that:
\begin{equation}
\Xi_{\Qq_i} \Xi_{\Qq_j} =  \Xi_{\Qq_j} \Xi_{\Qq_i}, \quad \forall \, i,j = 1,\ldots,2g. 
\end{equation}

\vspace{0.2cm}

In Fig.~\ref{Fig:Triangulation5}(b), we overlapped the $\mathfrak C$ and $\mathfrak Q$ topological contours of the torus, to highlight additional particularities:
\begin{itemize}

\item On one hand, $\Cc_1$ and $\Qq_1$ as well as $\Cc_2$ and $\Qq_2$  intersect at an odd number of fermion sites, hence $|\Cc_i \cap \Qq_i|$ = odd, $i=1,2$.

\item On the other hand, $\Cc_1$ and $\Qq_2$ as well as $\Cc_2$ and $\Qq_1$  intersect at an even number of fermion sites, hence $|\Cc_i \cap \Qq_j|$ = even, $i,j=1,2$, $i \neq j$.

\end{itemize}
By examining Fig.~\ref{Fig:GenusGG}, we can make very simple choices such that the above  characteristics persists for triangulations of higher genus surfaces. It then follows from \eqref{Eq:Xi2} that:
\begin{equation}
\Xi_{\Qq_i} S_{\Cc_j} = (-1)^{\delta_{ij}} S_{\Cc_j} \Xi_{\Qq_i}, \quad \forall \, i,j=1,\ldots,2g. 
\end{equation}
Hence, point iii) of Theorem~\ref{Th:Main} follows if we choose $\Xi^i = \Xi_{\Qq_i}\Pp$, $i=1,\ldots,2g$.

\section{Concluding Remarks}

The generic fermion models drawn from the sub-algebra $C^\ast(\Gamma_T,S_V; \, T,S\in \Ll)$ do not conserve the particle numbers, in general, and we have not been yet able to construct a model that does conserve the fermion number. The issue remains opened for now.

\vspace{0.2cm}

Since $C^\ast(\Gamma_T,S_V; \, T,S\in \Ll)$ is non-commutative, we were also un-able to construct an exactly solvable model. At the moment, we are investigating our predictions numerically, using the methods described in \cite{ProdanArxiv2019}.

\vspace{0.2cm}

We insisted in presenting an alternative proof of the topological spectral degeneracy that relies on finding an embedding of $\pi_1^b(M)$ rather than of a projective representation of $\ZM_2^{\times 4 g}$  because the existence of the $\Xi$'s, in such a simple form, appears to us more like a miracle. It might be the case that in other contexts, one will not be able to find the $\Xi$'s, yet the topological degeneracy can be entirely inferred from the algebra of $\Sigma$'s.


\section{References}

\begin{thebibliography}{9}

\bibitem{ArmstrongBook} M. A. Armstrong, {\sl Basic topology}, (Springer, Berlin, 1983).

\bibitem{BratelliBook2} O. Bratteli, D. W. Robinson, {\sl Operator algebras and quantum statistical mechanics 2}, (Springer, Berlin, 2002).

\bibitem{CarforaBook} M. Carfora, A. Marzuoli, {\sl Quantum Triangulations}, (Springer, Berlin, 2012).

\bibitem{ChenAOP2018} Y.-A. Chen, A. Kapustin, D. Radičevićb, {\sl Exact bosonization in two spatial dimensions and a new class of lattice gauge theories}, Annals of Physics {\bf 393}, 234-253 (2018).


Author links open overlay panel


\bibitem{ColemanBook} P. Coleman, {\sl Introduction to many-body physics}, (Cambridge Univ. Press, Cambridge, 2015).

\bibitem{DavidsonBook} K. R. Davidson, {\sl $C^*$-Algebras by Example}, (AMS, Providence, 1996).


\bibitem{DixmierBook1} J.~Dixmier, {\sl $C^\ast$-algebras}, (North-Holland Publishing, New York, 1977).

\bibitem{EmmanouilBook} I. Emmanouil, {Idempotent matrices over complex group algebras}, (Springer, Berlin, 2006).

\bibitem{Giol} J. Giol, {\sl Similarity implies homotopy of idempotents in Banach algebras of stable rank one}, Arch. Math. {\bf 88}, 235-238 (2007).

\bibitem{GuPRB2014} Z.-C. Gu, Z. Wang, X.-G. Wen, {\sl Lattice Model for Fermionic Toric Code}, Phys. Rev. B {\bf 90}, 085140 (2014).

\bibitem{JonesBook1} V. F. R. Jones, {\sl Subfactors and knots}, (AMS, Providence, 1991).

\bibitem{KitaevAOP2003} A. Y. Kitaev, {\sl Fault-tolerant quantum computation by anyons}, Annals of Physics {\bf 303}, 2-30 (2003).

\bibitem{KitaevAOP2006} A. Y. Kitaev, {\sl Anyons in an exactly solved model and beyond}, Annals of Physics {\bf 321}, 2–111 (2006).

\bibitem{KitaevOxford2008} A. Kitaev, C. Laumann, {\sl Topological phases and quantum computation},  in: ``Exact methods in low-dimensional statistical physics and quantum computing,'' Lecture Notes of the Les Houches Summer School No.89. Oxford University Press , Oxford, pp. 101-125 (2008).

\bibitem{ProdanArxiv2019} E. Prodan, {\sl Computational Many-Body Physics via 
$\Mm_{2^q}$ Algebra},  arXiv:1906.07309 (2019).

\bibitem{ShuklaPRB2020} Sujeet K Shukla, Tyler D Ellison, Lukasz Fidkowski, {\sl Tensor network approach to two dimensional bosonization}, Phys. Rev. B {\bf 101}, 155105 (2020).

\bibitem{RenaultBook} J. Renault, {\sl A groupoid approach to $C^\ast$-algebras}, (Springer-Verlag, Berlin, 1980). 

\bibitem{WenBook1} X.-.G. Wen, {\sl Quantum field theory of many-body systems}, (Oxford University Press, Oxford, 2004).

\bibitem{Wene} G. P. Wene, {\sl The idempotent structure of an infinite dimensional Clifford algebra},  in: A. Micali, R. Boudet, J. Helmstetter (eds) {\sl Clifford Algebras and their Applications in Mathematical Physics}, Fundamental Theories of Physics {\bf 47}, Springer, Dordrecht, (1992). 
 
\bibitem{WenBook2} B. Zeng, X. Chen, D.-L. Zhou, X.-Ga. Wen, {\sl Quantum Information Meets Quantum Matter},  (Springer, Berlin, 2019).


\end{thebibliography}
\end{document}